\documentclass[fleqn,usenatbib]{mnras}

\usepackage{newtxtext,newtxmath}

\usepackage[T1]{fontenc}
\usepackage{ae,aecompl}

\usepackage{graphicx}	
\usepackage{amsmath}	
\usepackage{amssymb}	

\usepackage{lipsum} 
 
   \title[The effects of surface fossil magnetic fields on massive star evolution: I.]{The effects of surface fossil magnetic fields on massive star evolution: I. Magnetic field evolution, mass-loss quenching and magnetic braking}

\author[Z. Keszthelyi et al.]{
Z. Keszthelyi$^{1,2}$\thanks{E-mail: zsolt.keszthelyi@rmc.ca},
G. Meynet$^{3}$,
C. Georgy$^{3}$,
G.A. Wade$^{1}$,
V. Petit$^{4}$,
A. David-Uraz$^{4}$
\\
$^{1}$Department of Physics and Space Science, Royal Military College of Canada, PO Box 17000 Station Forces, Kingston, \\ ON, K7K~0C6, Canada\\
$^{2}$Department of Physics, Engineering Physics and Astronomy, Queen's University, 99 University Avenue, Kingston, \\ ON, K7L~3N6, Canada  \\
$^{3}$Geneva Observatory, University of Geneva, Maillettes 51, 1290 Sauverny, Switzerland \\
$^{4}$Department of Physics and Astronomy, University of Delaware, 217 Sharp Lab, Newark, DE 19716, USA
}

\date{Accepted XXX. Received YYY; in original form ZZZ}

\pubyear{2019}

\begin{document}
\label{firstpage}
\pagerange{\pageref{firstpage}--\pageref{lastpage}}
\maketitle

\begin{abstract}
Surface magnetic fields have a strong impact on stellar mass loss and rotation and, as a consequence, on the evolution of massive stars. In this work we study the influence of an evolving dipolar surface fossil magnetic field with an initial field strength of 4 kG on the characteristics of 15 M$_{\odot}$ solar metallicity models using the Geneva stellar evolution code. Non-rotating and rotating models considering two different scenarios for internal angular momentum transport are computed, including magnetic field evolution, mass-loss quenching, and magnetic braking. Magnetic field evolution results in weakening the initially strong magnetic field, however, in our models an observable magnetic field is still maintained as the star evolves towards the red supergiant phase. At the given initial mass of the models, mass-loss quenching is modest. Magnetic braking greatly enhances chemical element mixing if radial differential rotation is allowed for, on the other hand, the inclusion of surface magnetic fields yields a lower surface enrichment in the case of near solid-body rotation. Models including surface magnetic fields show notably different trends on the Hunter diagram (plotting nitrogen abundance vs $v \sin i$) compared to those that do not. The magnetic models agree qualitatively with the anomalous `Group 2 stars', showing slow surface rotation and high surface nitrogen enhancement on the main sequence.
\end{abstract}




\section{Introduction}
Extensive spectropolarimetric surveys have revealed that $7\% $ of massive OB stars host an observable surface magnetic field \citep{morel2015,fossati2015,wade2016,grunhut2017,shultz2018}. These fields are strong (on the order of kG), mostly dipolar, stable on long time-scales, and are understood to be of fossil origin \citep{donati2009}.    

Surface magnetic fields exhibit a complex interaction with the stellar winds of hot stars, channelling the outflow and confining a significant fraction of the wind in a closed, co-rotating magnetosphere \citep{babel1997,ud2002,townsend2005,petit2013,bard2016,owocki2016}. Two key `surface' phenomena resulting from this interaction and important in the evolutionary context have been identified: \textit{magnetic braking} (increasing the angular momentum loss of the star) and \textit{mass-loss quenching} (decreasing the mass-loss rate of the star). Simple prescriptions, taking into account magnetic, wind, and rotational properties, have been developed and can be exploited to quantify these phenomena \citep{ud2008,ud2009}.

First steps towards exploring the evolutionary consequences of surface fossil magnetic fields of massive stars have previously been taken by a few investigations. 

\cite{mahes1994} studied the evolution of rotation and surface fossil magnetic fields in 60 M$_{\odot}$ models using a kinematic approach and placed constraints on the Be and Wolf-Rayet phases. However, the stellar evolution model was not modified to account for rotation or magnetic fields.
  
\newpage
\cite{meynet2011} studied magnetic braking in massive star models. The key results of their study are that i) magnetic braking can greatly deplete the core angular momentum reservoir on the main sequence, and ii) a surface magnetic field can significantly modify the chemical enrichment of the star. However, a basic limitation of the study was the assumption that the surface magnetic field strength remained constant during stellar evolution (thus the magnetic flux increased) and the fact that mass-loss quenching by the magnetic field was not considered.

Recently, \cite{petit2017} studied mass-loss quenching alone in non-rotating models with evolving surface field strength using the \textsc{mesa} code \citep{paxton2013}, and \cite{georgy2017} presented first results including both mass-loss quenching and magnetic braking in the Geneva code \citep{eggenberger2008}. \cite{petit2017} and \cite{georgy2017} showed that mass-loss quenching allows the star to maintain a higher mass during its evolution, respectively in the context of `heavy' stellar-mass black holes such as those whose merger was reported by aLIGO (\citealt{abbott2016,abbott2017a,abbott2017b}), and in the context of pair-instability supernovae. Hence, these studies have shown that magnetic massive stars even at solar metallicity could serve as progenitors in both cases.

Since both the studies by \cite{petit2017} and \cite{georgy2017} focused primarily on mass-loss quenching, massive and very massive stars (40 - 80 M$_\odot$ and 200 M$_\odot$, respectively) were modelled - objects that are considerably rare in nature and representative of only a small fraction of known magnetic hot stars.

\cite{potter2012b} studied magnetic braking for a population of stars using the \textsc{rose} code \citep{potter2012a}. They implemented an $\alpha$-$\Omega$ radiative dynamo mechanism to induce and maintain an internal large-scale magnetic field, while magnetic braking was attributed to an external surface field. A key result of their study is that they obtain a population of slowly rotating, nitrogen-enriched stars with surface magnetic field strengths that could potentially be detected by current instrumentation; however, this modeled population is strictly limited to a mass range close to 10 M$_\odot$. Furthermore, \cite{quentin2018} studied the evolution of a typical intermediate-mass star, considering a large-scale, dynamo-generated internal magnetic field. The magnetic field equations are coupled to stellar rotation, and the magneto-rotational instability is considered (see also \citealt{wheeler2015}). 
While currently the fossil field hypothesis is favoured to explain the origin of observed magnetism at the surface of hot stars, these approaches - in addition to studies that investigated the Spruit-Tayler \citep{spruit2002,tayler73} dynamo mechanism \citep{maeder2003,maeder2004,maeder2005,heger2005} - may also prove instructive for future studies, especially to account for angular momentum redistribution by fossil fields.
 
Considering a different approach, \cite{petermann2015} computed stellar evolution models implementing a reduction in the convective core size caused by a strong fossil magnetic field. While this reduction is accomplished by an arbitrary parameter, the physical nature of magnetic fields suppressing convection has been studied in a variety of contexts, e.g., subsurface convection zones of massive stars \citep{sundqvist2013}, Ap stars \citep{michaud1970}, late-type main-sequence stars \citep{cox1981,chabrier2007} and the atmospheres of white dwarfs \citep{tremblay2015}. 

In addition to these studies of single-star evolution, recent studies by \cite{chen2016} and \cite{song2018} also consider surface magnetic fields, however in the context of intermediate-mass and massive binary evolution.  

Apart from these studies, surface fossil magnetic fields are usually not considered in massive star evolution models \citep{brott2011,ekstroem2012,paxton2013}, although it has been speculated that surface magnetism may play a significant role in resolving several problems. For instance, some anomalous trends on the Hunter diagram \citep{hunter2008} showing nitrogen abundance as a function of projected rotational velocity have been speculated to be caused by stellar magnetism \citep{hunter2009,brott2011b}. We will discuss this in detail in Section~\ref{sec:hunt}. 

The important evolutionary consequences of magnetic fields, established by previous studies, suggest that the application of non-magnetic stellar evolution models to magnetic hot stars may be inappropriate, and may result in erroneous conclusions when deriving stellar parameters from non-magnetic models. The present study seeks to address two important questions related to this uncertainty:
\begin{itemize}
\item When comparing known magnetic stars to the predictions of stellar evolution models, can non-magnetic models yield reasonable stellar parameters?
\item To what extent can the newly-incorporated prescription of an evolving surface magnetic field inform and potentially account for debated problems in stellar astrophysics, for instance, for the surface nitrogen enrichment of massive stars?  
\end{itemize}

The scope and motivation of this study is to assess the cumulative influence of magnetic mass-loss quenching, magnetic braking and magnetic field evolution, each of these components having heretofore been considered separately. Moreover, by predicting the evolution of surface field strengths beyond the main sequence, this study provides benchmarks for the detection of magnetic fields on hot, evolved stars. 

To this extent we perform model calculations with the Geneva code for a typical massive star model with initial mass of $M_{\star,\rm ini} = 15$ M$_{\odot}$ at solar ($Z_{\rm ini} = 0.014$, \citealt{asplund2005,ekstroem2012}) metallicity, and investigate how the fossil field changes with time, assuming that the surface magnetic flux is conserved, and no other significant flux decay or enhancement processes occur.

This work is structured as follows: In Section 2 we describe the model including the evolution of a surface magnetic field. In Section 3, we compare the evolution of non-magnetic and magnetic models. In Section 4 we discuss the rotational and angular momentum evolution, the nitrogen abundance vs the rotational velocity, and the evolution of the surface magnetic field and related magnetospheric parameters. In Section 5 we draw conclusions regarding surface magnetism in massive stars and relevant points for future surveys.


\section{Methods}\label{sec:met}

\subsection{Surface fossil magnetic field prescription}

We use the Geneva stellar evolution code (\textsc{genec}, \citealt{eggenberger2008}) for our model calculations. We consider that the  magnetic flux is frozen into the plasma (Alfv\'en's theorem, \citealt{alfven1942}) thereby leading to the local conservation of magnetic flux over time. Thus we assume that the change in surface magnetic field strength is related to the change in the stellar radius as:
\begin{equation}\label{eq:field}
B_p  = B_{\rm p, ini}  \left( \frac{R_{\rm \star,  ini}}{R_\star } \right)^{2} \, , 
\end{equation}
where $B_p$ is the polar magnetic field strength at the stellar surface (photosphere), $B_{\rm p, ini}$ is the initial, zero age main sequence (ZAMS) surface polar magnetic field strength, $R_\star $ is the stellar radius, and $R_{\rm \star, ini}$ is the ZAMS radius\footnote{There is an important difference between the spatial and the time dependence of the magnetic field. While flux conservation yields a $B_p (t) \propto R_{\star}^{-2}(t)$ time dependence, the spatial variation in the radial direction for a dipole field configuration is described by $B (r) \propto r^{-3}$.}. 
$B_p$ and $R_{\star}$ evolve in concert with the star, that is $B_p = B_p (t)$ and $R_{\star} = R_{\star} (t)$, however for simplicity we will not denote explicitly the time dependence since it is understood for all but the initial quantities. 

We account for mass-loss quenching and magnetic braking due to a fossil magnetic field as previously included in stellar evolution model calculations by \citet{meynet2011}, \citet{keszthelyi2017b}, \citet{petit2017}, \citet{georgy2017}, and \citet{song2018}. We employ the prescriptions and formalism based on the results of \citet{ud2002}, \citet{ud2008,ud2009}, and \citet{petit2013}. 

Mass-loss quenching results from stellar wind plasma being trapped by closed magnetic field lines near the stellar magnetic equator, causing cooled gas to return to the stellar surface \citep{ud2002}. It leads to a fractional reduction $f_B$ in the overall mass-loss rate, also defined as the escaping wind fraction, 
\begin{equation}\label{eq:fb} 
f_B \approx 1 - \sqrt{ 1 - \frac{R_\star}{R_c}  }  \, , 
\end{equation}
where $R_c$ is the closure radius \citep{ud2008} that defines the distance of the last closed magnetic loop from the stellar surface, and can be expressed in terms of the Alfv\'en radius $R_A$ (see below) as:
\begin{equation}
R_c \approx R_\star + 0.7 ( R_A - R_\star ) \, .
\end{equation}
Thus the effective mass-loss rate becomes:
\begin{equation}\label{eq:quench}
\dot{M} = f_B \cdot \dot{M}_{B=0} \, ,
\end{equation}
where $\dot{M}_{B = 0}$ is the mass-loss rate the star would have in the absence of a magnetic field. It should be noted that Equation~\ref{eq:fb} does not take into account the latitudinal variation in the mass flux due to the tilt of the magnetic field with respect to the normal to the surface, and that this approximation is particularly good for large values of $R_c$, but somewhat less robust otherwise. There are, however, leakage mechanisms that may allow for mass to escape from closed loops if the star is rotating rapidly, but this is typically only important for the early evolution as the star quickly spins down \citep{owocki2018}. Thus, Equation \ref{eq:fb} is indeed an approximation for the reduction in mass-loss rates on average. Furthermore, it should be noted that when the stellar radius is smaller, that is the star is more compact, the escaping wind fraction becomes larger.
This formulation in \textsc{genec} is coupled to the stellar wind calculations, namely any adopted mass-loss scheme can be applied to obtain $\dot{M}_{B=0}$, and then this value is scaled by $f_B$ to account for the reduction by the magnetic field.

Magnetic braking accounts for the removal of an additional amount of angular momentum, over and above that naturally removed by (non-magnetic) mass loss, due principally to Maxwell stresses:  
\begin{equation}\label{eq:br}
\dot{J}  = \frac{2}{3} \dot{M}_{B = 0} \Omega_\star R_\star^2 \, R_A^2 = \frac{2}{3} \dot{M}_{B = 0} \Omega_\star R_\star^2 \, \left[ (\eta_\star + 0.25)^{0.25} + 0.29   \right]^2  \, , 
\end{equation}
where $\Omega_\star$ is the surface angular velocity, $R_A$ is the Alfv\'en radius for a dipole configuration, and $\eta_\star$ is the equatorial wind magnetic confinement parameter \citep{ud2009}, defined as 
\begin{equation}\label{eq:eta}
\eta_\star = \frac{B_p^2R_\star^2}{4 \dot{M}_{B=0} v_{\infty}} \, , 
\end{equation}
where $v_{\infty}$ is the terminal wind velocity. 
For consistency, we also updated the Geneva code to systematically calculate $v_\infty$ from the escape velocity $v_{\rm esc}$ (following \citealt{kudritzki2000} and \citealt{vink2001}) as $v_\infty =~2.6 v_{\rm esc}$ for $T_{\rm eff} > 20$~kK,  $v_\infty =~1.3 v_{\rm esc}$ for 20~kK$~>~ T_{\rm eff} >~10$~kK, and $v_\infty = 0.7 v_{\rm esc}$ for $T_{\rm eff} <~10$ kK. 

Since magnetic braking modifies the angular velocity of the stellar surface, the Geneva code implements Equation \ref{eq:br} as a boundary condition of the internal angular momentum transport equation at the stellar surface, and modifies the total angular momentum content of the star.

\subsection{Internal angular momentum transport}
Possibly one of the most important questions regarding stellar rotation is whether the internal layers of a star could be considered to achieve solid-body rotation or radial differential rotation. In the former case, some mechanism must couple the stellar core and envelope.

This question divides our model calculations into two branches to account for this uncertainty. Although solid-body rotation could be enforced artificially, for main sequence models it is also a known natural consequence of adopting the viscosity of a Spruit-Tayler (ST, \citealt{spruit2002,tayler73}) dynamo mechanism in the radiative envelope of the star to account for the angular momentum transport \citep{maeder2003,maeder2004,maeder2005,heger2005,petrovic2005,brott2011,keszthelyi2017a,song2016}. Therefore, two scenarios are considered regarding the angular momentum transport, which, following \cite{zahn92} and \cite{chaboyer1992}, can be written as:
\begin{equation}\label{eq:gam}
 \rho \frac{ \partial (r^2 \Omega)  }{ \partial t}    = \frac{1}{5 r^2} \frac{\partial}{\partial r } \left( \rho r^4 \Omega U(r)  \right) + \frac{1}{r^2} 
 \frac{\partial}{\partial r} \left( \rho  r^4  \, D_{\rm AM} \,  \frac{\partial \Omega}{\partial r}   \right) \, , 
\end{equation}
where $\rho$ is the density, $r$ is the radius, $t$ is time, $\Omega$ is the angular velocity in a given layer of the star, $U(r)$ is the amplitude of the vertical component of the meridional circulation velocity, and $D_{\rm AM}$ is the sum of diffusion coefficients that contribute to angular momentum transport. The first term on the r.h.s. of Equation~\ref{eq:gam} is the advective term, while the second term is the diffusive term.

We consider two approaches to construct $D_{\rm AM}$ in radiative zones. In the differentially rotating case, when the transport is less efficient, 
\begin{equation}
D_{\rm AM,1} = D_{\rm shear}  \, , 
\end{equation}
where $D_{\rm shear}$ is the diffusion coefficient due to shears. In this case, both advective and diffusive terms are used in Equation~\ref{eq:gam} and the meridional currents (\citealt{eddington1925,sweet1950}) are thus accounted for via advection (see \citealt{zahn92,maeder1998,meynet2013}). 

In the solid-body rotating case, when the transport is more efficient, 
\begin{equation}
D_{\rm AM,2} = D_{\rm shear} + D_{\rm circ,H} + D_{\rm ST} \, , 
\end{equation}
where $D_{\rm circ,H}$ is a diffusion coefficient accounting for meridional circulation and $D_{\rm ST}$ is a diffusion coefficient resulting from the viscosity of the ST dynamo mechanism (Equation~15 of \citealt{maeder2005}). In this case, the advective term in Equation \ref{eq:gam} can be neglected, justifying a purely diffusive treatment \citep{song2016}.

In the following, by \textit{magnetic models} we always refer to models with surface fossil magnetic fields. We stress here that the ST dynamo is only introduced in order to achieve a flat angular velocity profile on the main sequence since it results in a large diffusion coefficient for angular momentum transport\footnote{Strictly speaking, the inclusion of $D_{\rm ST}$ does not achieve completely perfect solid-body rotation since a minimal shear is required to operate the ST mechanism, which in turn significantly flattens the $\Omega$ profile and enhances meridional circulation \citep{maeder2005}. Moreover, this mechanism is expected to weaken somewhat after the main sequence, although still maintaining a nearly flat $\Omega$ profile.}. Including or neglecting the viscosity of the hypothetical ST dynamo in the equation of internal angular momentum transport serves only to establish two fundamentally different rotational configurations which are of interest to study.

%
%
%
\begin{table}
\caption{Geneva stellar evolution models computed in this study. In the brackets we denote rotation `R', diffusion coefficient from Spruit-Tayler dynamo `S', and surface magnetic fields `B', for clarity.}
\begin{tabular}{lccc}   
\hline
Model & $v_{\rm rot, ini}$  & $D_{\rm ST}$ & $B_{\rm p,ini}$ (surface) \\
 & [km s$^{-1}$] &  & [kG] \\
\hline \hline
M1 (- - -)& 0 & - & 0 \\
M2 (- - B) & 0 & - & 4 \\
M3 (R - -) & 200 & no &0 \\
M4 (R - B) & 200 & no &4  \\
M5 (R S -) & 200 & yes &0 \\
M6 (R S B) & 200 & yes &4 \\
\hline
\end{tabular} \label{tab:tm}
\end{table}
%
 
%
%
%
\subsection{General model parameters}

The models follow the conventional computation of a Geneva stellar evolution model \citep{ekstroem2012}. The most important initial parameters of the models are outlined in Table \ref{tab:tm}. An initial mass of $M_{\star,\rm ini} = 15$ M$_{\odot}$ and solar metallicity ($Z_{\rm ini} =~ 0.014$) are adopted. Models are computed from the zero age main sequence until they cross the Hertzsprung-Russell diagram (HRD) from the `blue' to the `red' part and start to deplete core helium. 
The initial elemental abundances follow the \cite{asplund2005} ratios, except for neon, which is adopted from \cite{cunha2006}. We compute rotating models with $v_{\rm rot, ini} = ~200 \, \mathrm{km \, s^{-1}}$ ($ v_{\rm rot, ini}/v_{\rm crit, ini} \sim 0.3$), as well as non-rotating models. The implementation of the rotating models follows the theory of \cite{zahn92} and \cite{maeder1998}. Core overshooting is calculated using the step method, extending the core by 10\% of the local pressure scale height. The core boundary is determined by the Schwarzschild criterion. Due to the initial mass of the model, we use the prescription of \cite{dej1988} to obtain the (non-magnetic) mass-loss rates. 
We adopt an initial surface magnetic field strength of 4 kG, considering that it is typical for magnetic massive OB stars based on the observed distribution of field strengths in young magnetic stars \citep{petit2013,shultz2018}.

For further details regarding \textsc{genec} and the details of the model calculations, see \cite{hirschi2004,ekstroem2012} and \cite{georgy2013}. 

Some of the models (those without surface magnetic fields) have already been computed with a previous version of the Geneva code. However, for consistency we recomputed these models. This means that from Table \ref{tab:tm}, models M1 (- - -) and M3 (R- -) resemble the non-rotating and rotating \citet{ekstroem2012} models, respectively, while M5 (RS-), including $D_{\rm ST}$ is similar to models presented by \cite{maeder2005}. In addition, \cite{meynet2011} computed models with surface magnetic fields (see Introduction), but those models considered magnetic braking under a constant magnetic field strength, and did not account for mass-loss quenching. In the models including surface magnetic fields M2 (- -B), M4 (R-B), and M6 (RSB) presented in this study, mass-loss quenching and magnetic braking are considered under self-consistently evolving surface magnetic parameters, obeying the magnetic field evolution model described by Equation \ref{eq:field}.
 

\section{Results}

\begin{figure*}
\includegraphics[width=6.5cm]{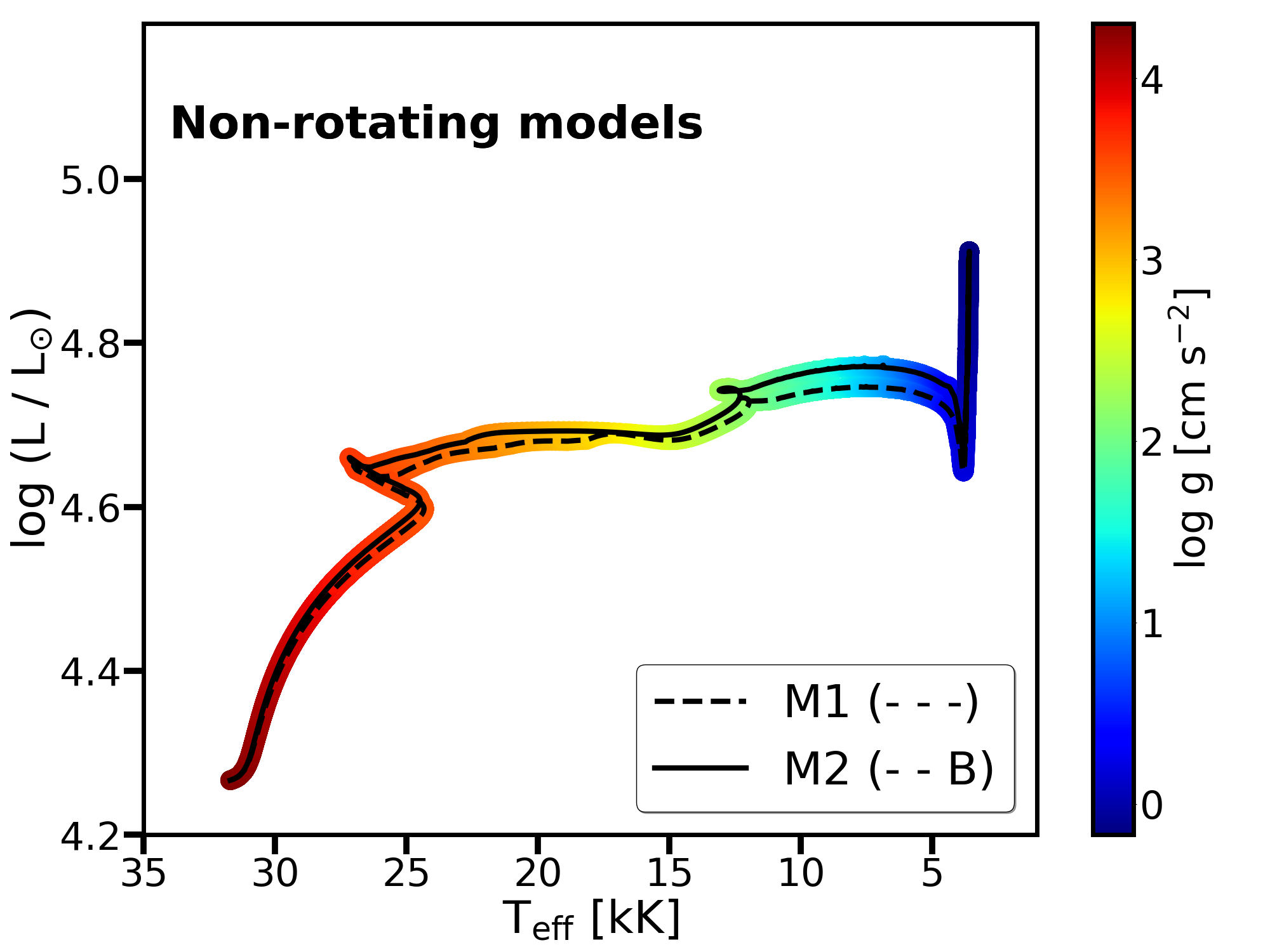}\includegraphics[width=6.5cm]{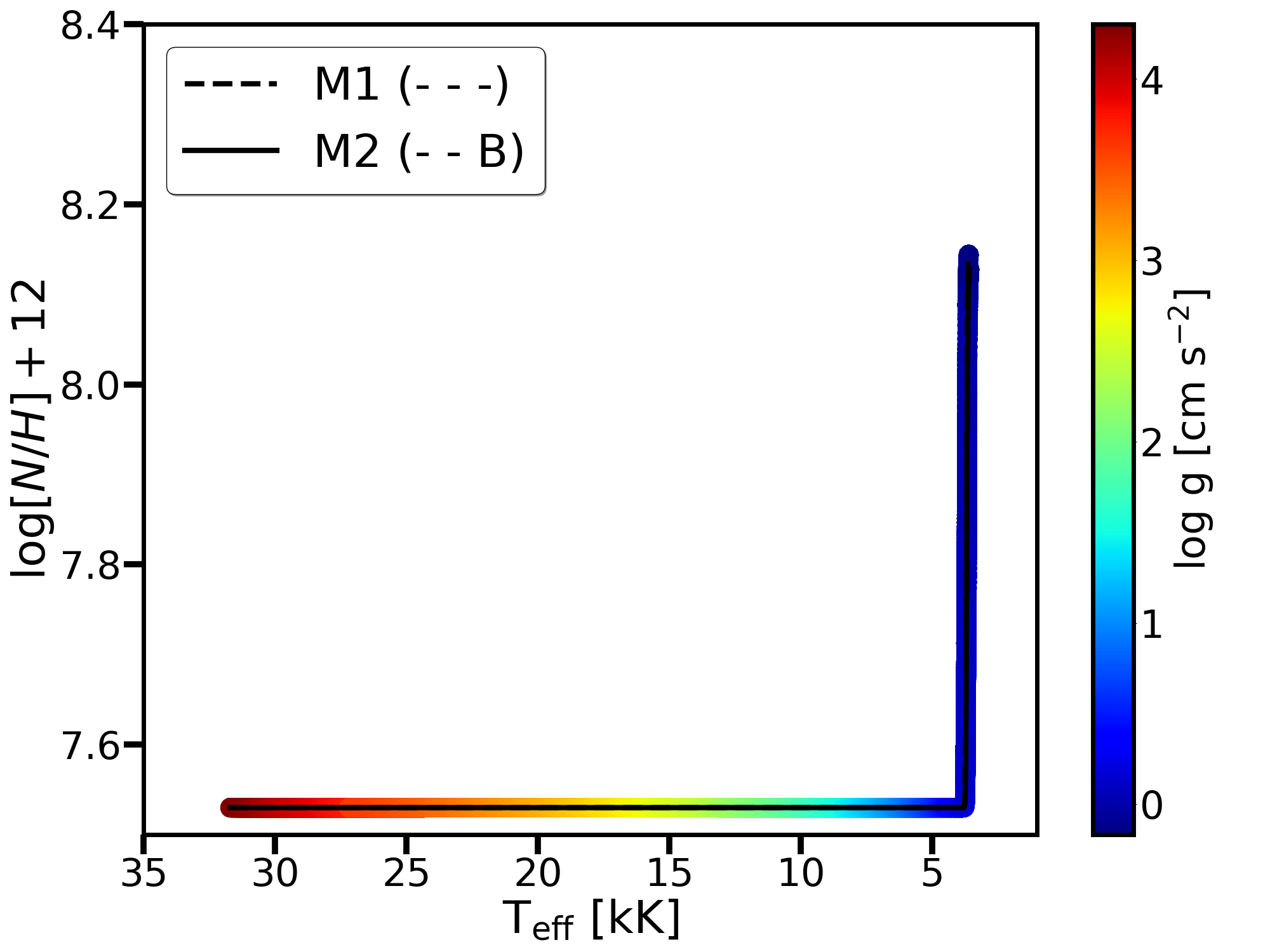}\includegraphics[width=5.5cm]{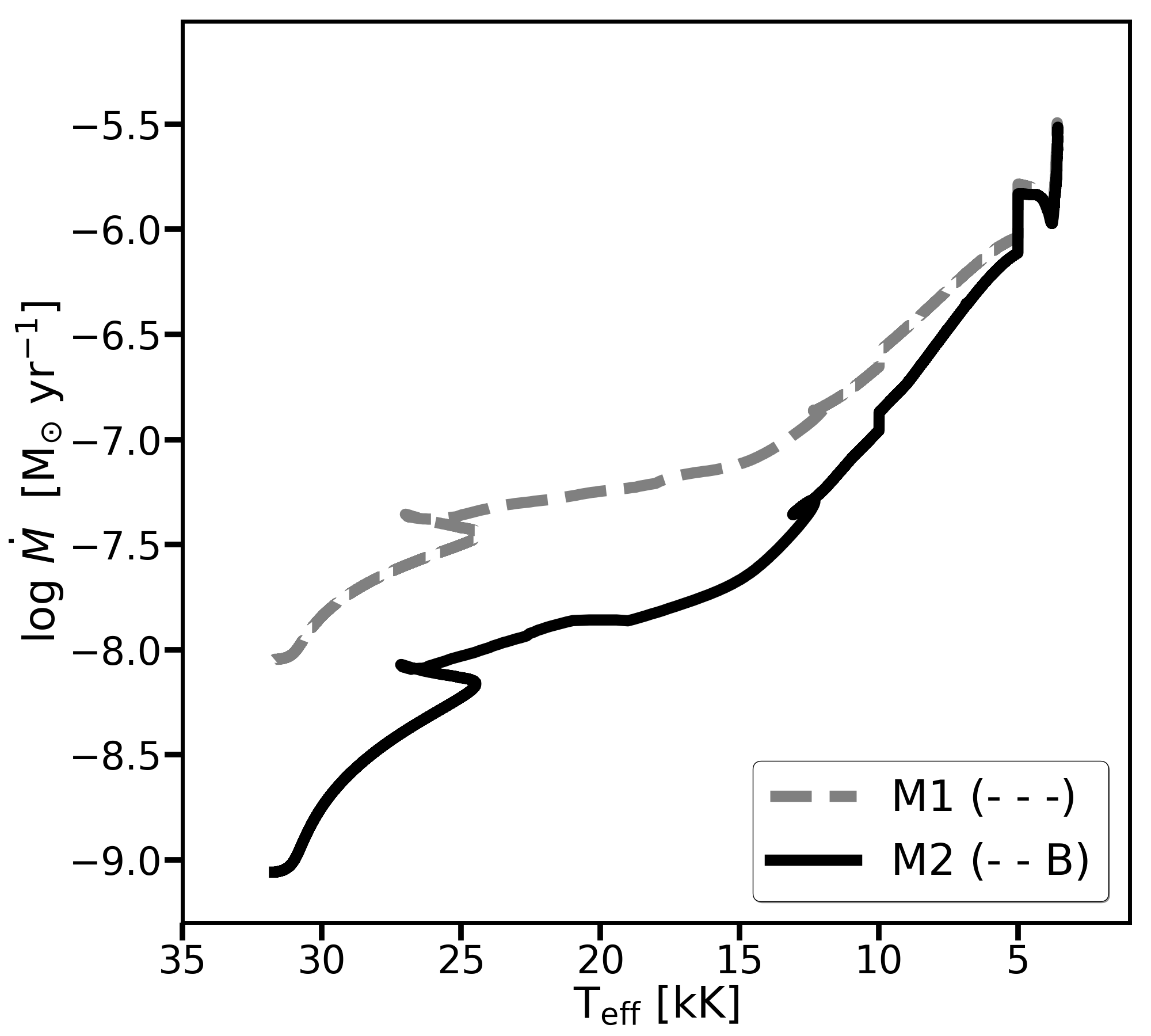} 
\caption{Comparison between the non-magnetic model M1 (- - -) and magnetic model M2 (- -B) without rotation on the HRD (left panel) colour-coded with the log of the surface gravity, their nitrogen enrichment (middle panel), and mass-loss history (right panel).} 
\label{fig:nrm1}
\end{figure*}

\begin{figure*}
\includegraphics[width=6.5cm]{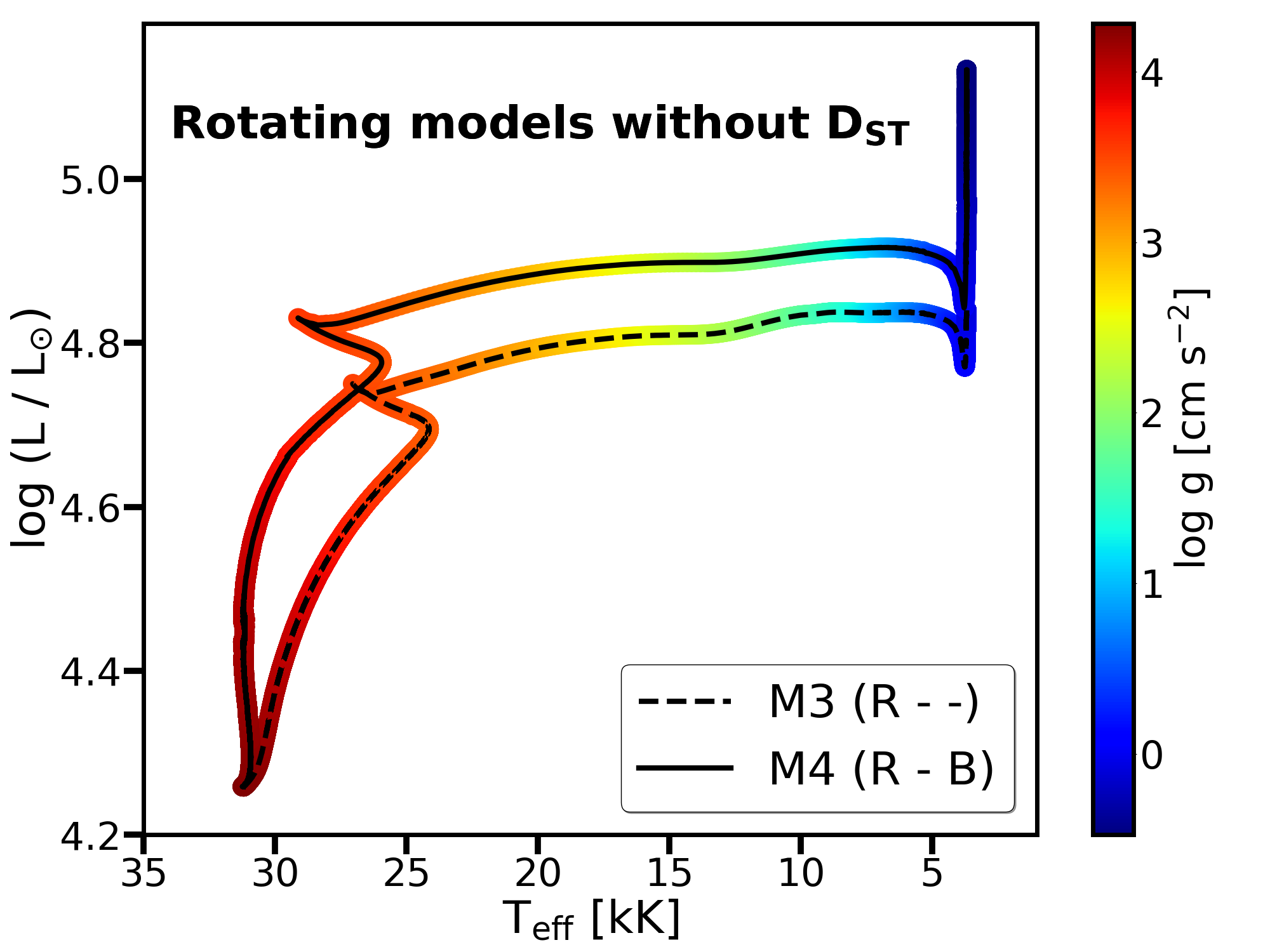}\includegraphics[width=6.5cm]{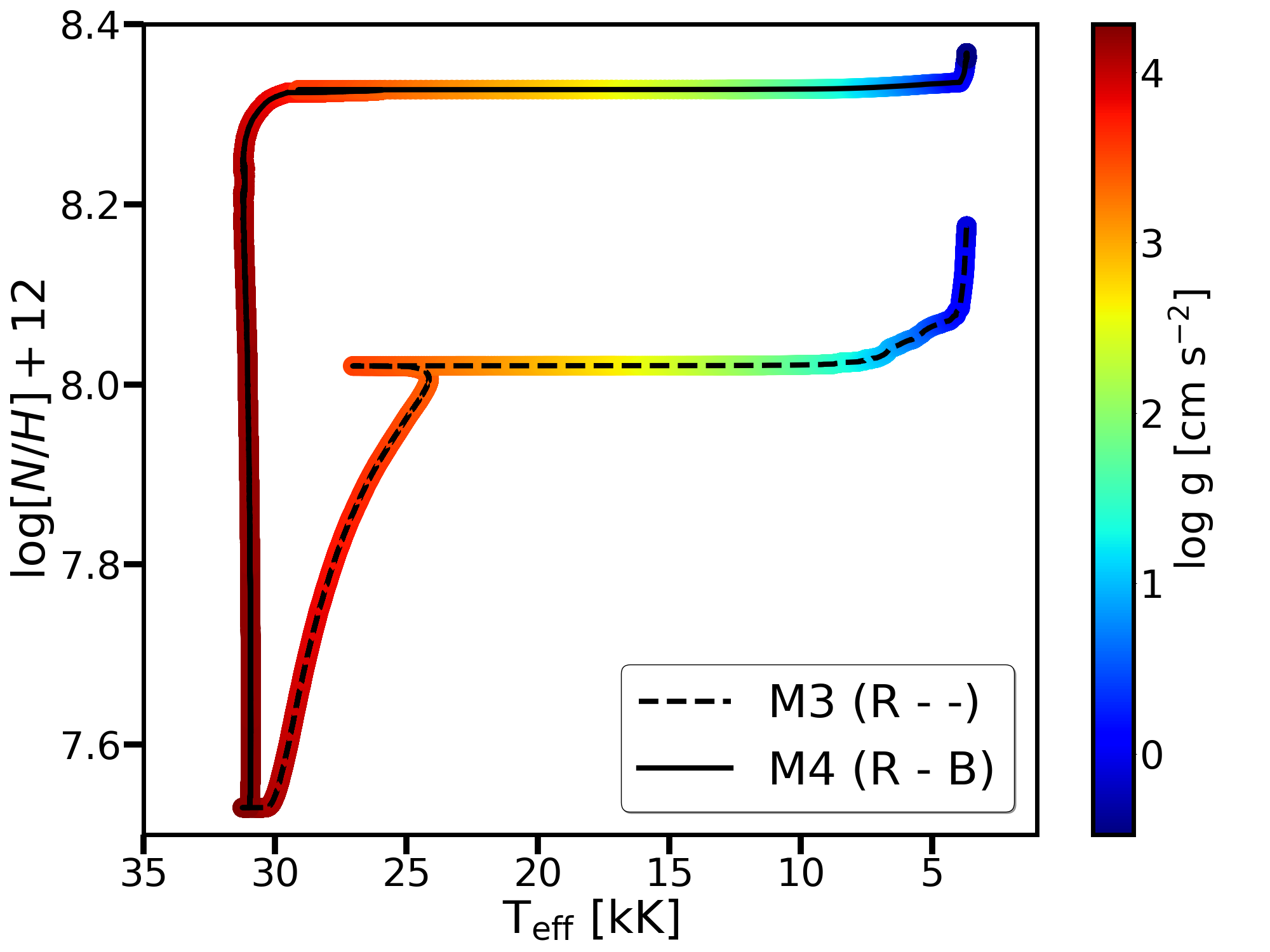}\includegraphics[width=5.5cm]{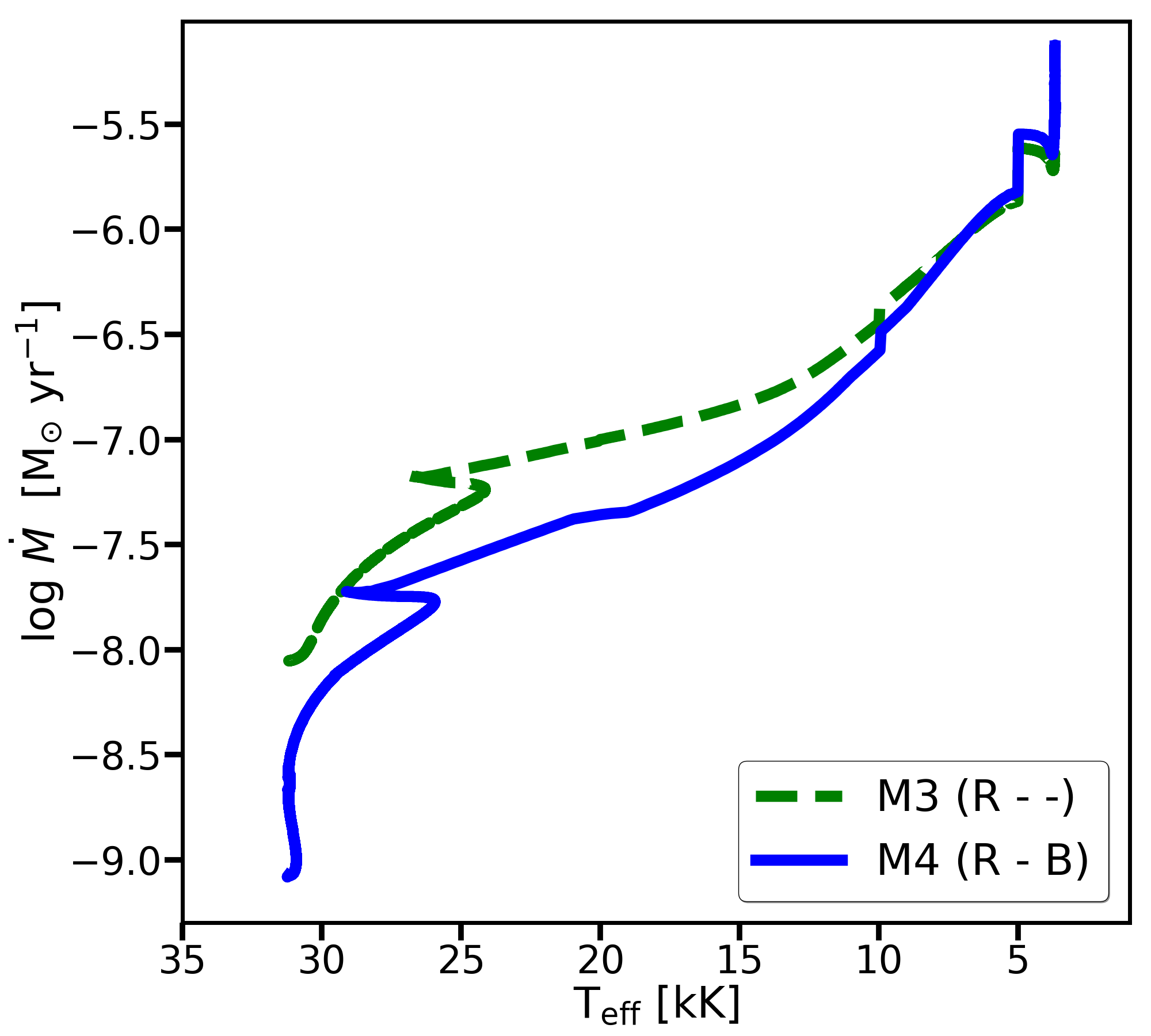} 
\caption{Same as Fig. \ref{fig:nrm1} but for models M3 (R- -) and M4 (R-B) with rotation and without D$_{\rm ST}$.} 
\label{fig:nrm2}
\end{figure*}

\begin{figure*}
\includegraphics[width=6.5cm]{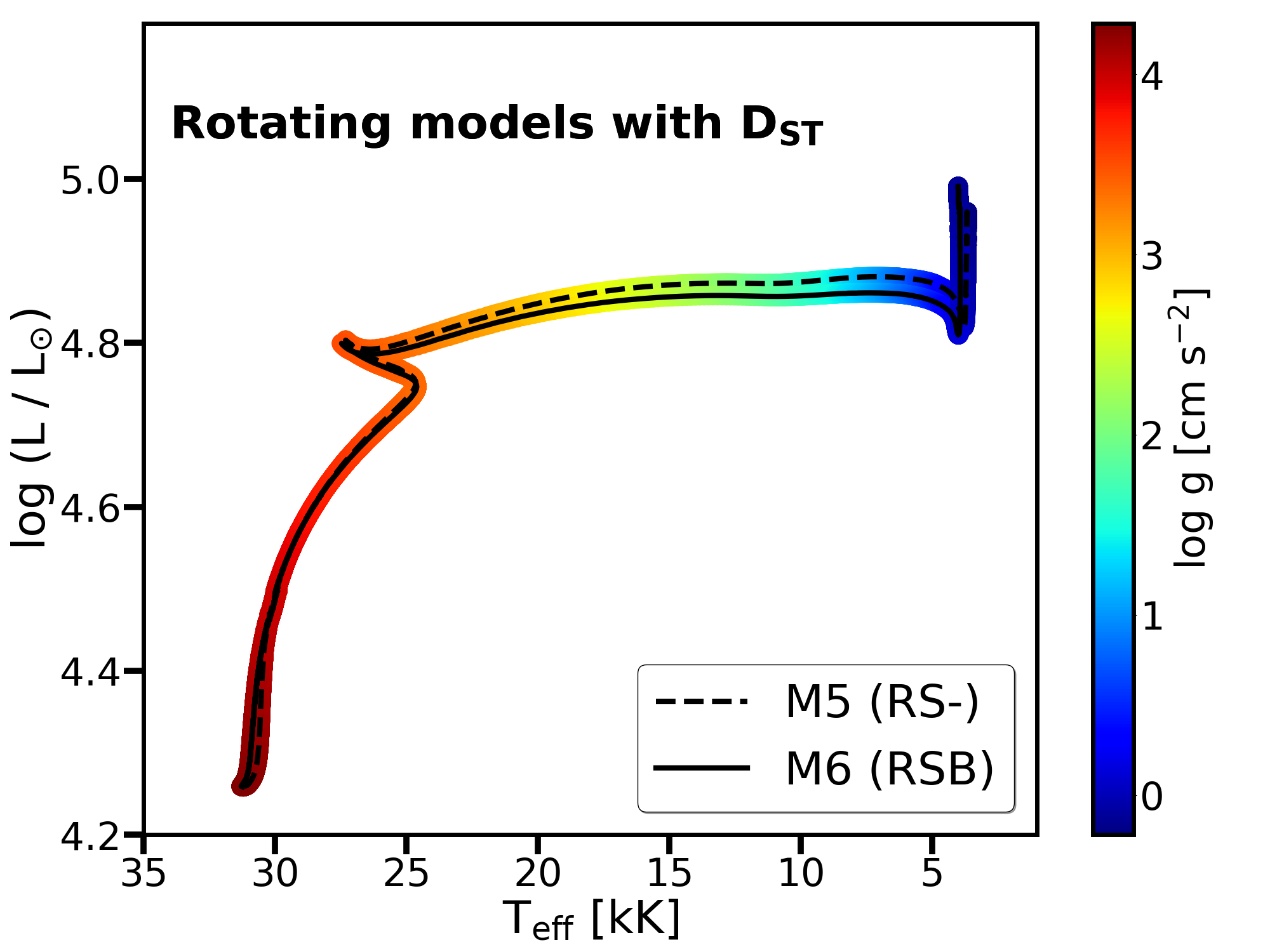}\includegraphics[width=6.5cm]{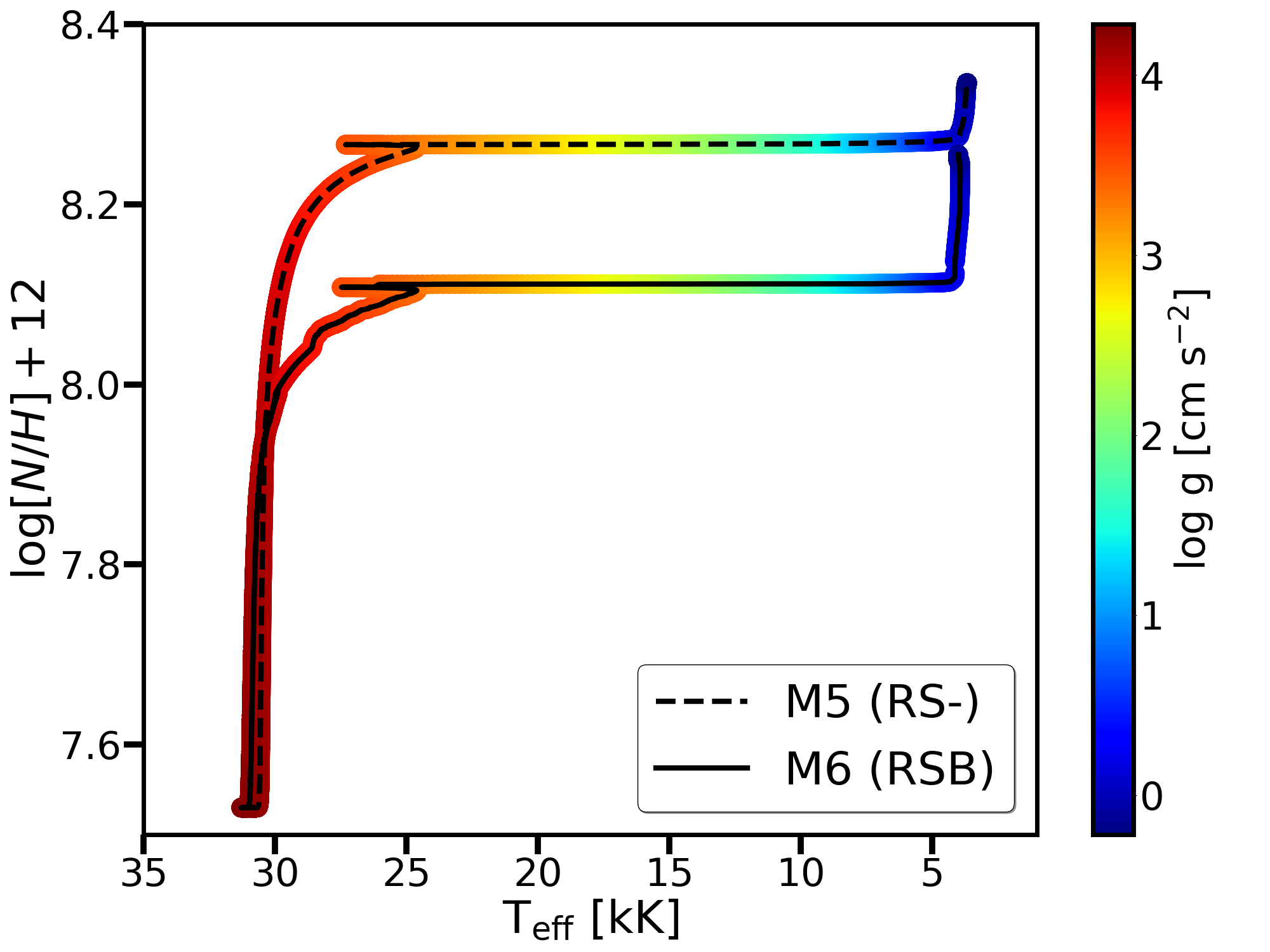}\includegraphics[width=5.5cm]{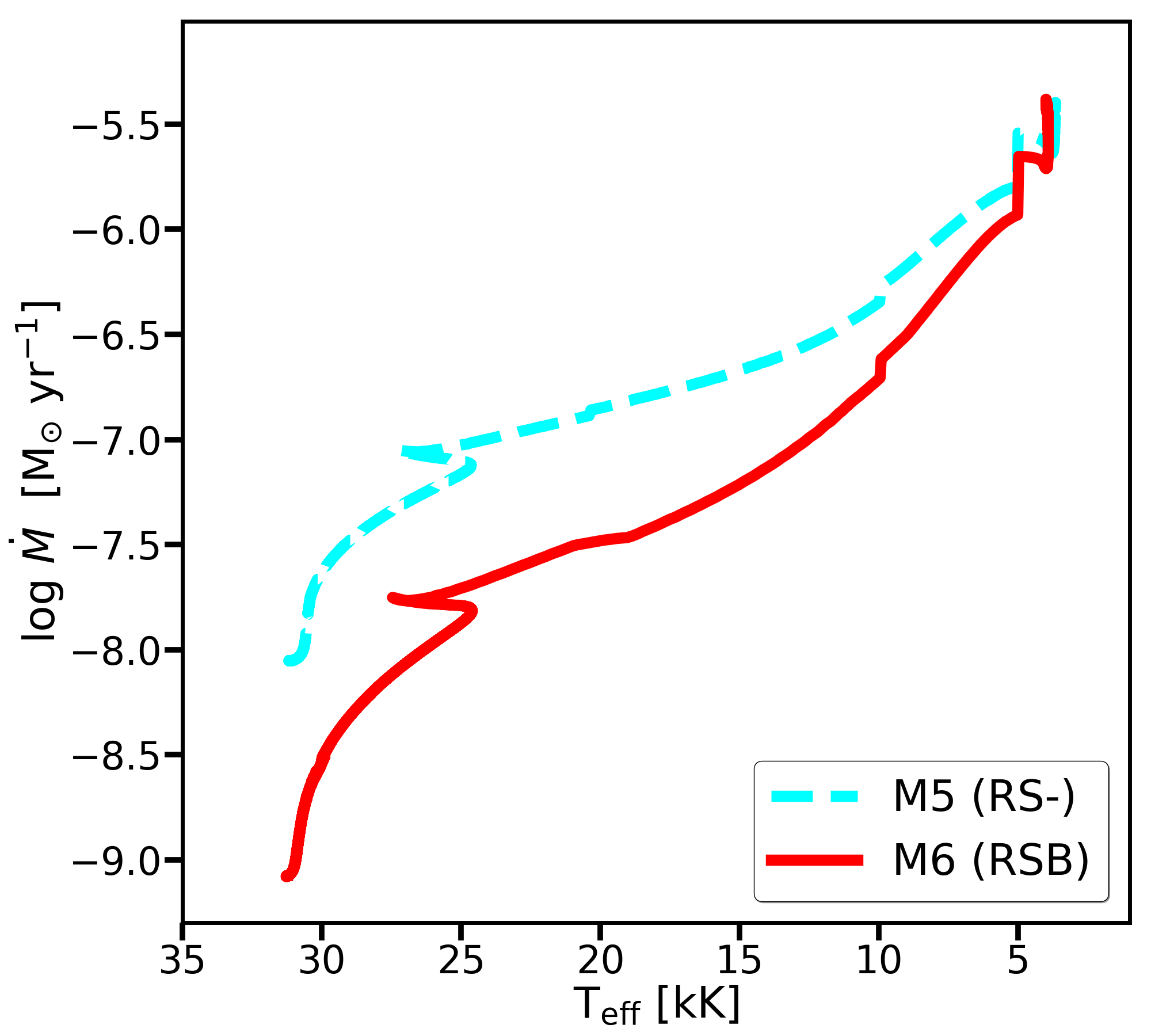}
\caption{Same as Fig. \ref{fig:nrm1} but for models M5 (RS-) and M6 (RSB) with rotation and with D$_{\rm ST}$.} 
\label{fig:nrm3}
\end{figure*}

In this section we present the results of our model calculations, and study the influence of individual model ingredients. We first compare a model without a surface magnetic field to a model with a surface magnetic field to quantify the impact of surface magnetism in stellar evolution models. This means that the discussion of the 6 computed models is divided into 3 subsections: 2 models without rotation, 2 models with rotation and without $D_{\rm ST}$, and 2 models with rotation and adopting $D_{\rm ST}$. Along these lines, Figures \ref{fig:nrm1}-\ref{fig:nrm3} display the model pairs on three panels: the HRD (left panels), the nitrogen abundance (middle panel), and mass-loss rates (right panels) vs the effective temperature.  
%
%
%
%

%
%
\subsection{Non-magnetic vs magnetic models without rotation: model M1 (- - -) and M2 (- -B)}

Figure \ref{fig:nrm1} compares the predictions of non-magnetic and magnetic models without rotation, M1 (- - -) and M2 (- -B), respectively. 

In the simplified case when rotation is neglected, the surface magnetic fields are only considered to decrease the overall mass-loss rate via mass-loss quenching. Such models were presented previously by \cite{petit2017}, but with significantly higher initial masses, until the models reached the terminal age main sequence (TAMS).

The mass loss by stellar winds in the case of 15 M$_\odot$ models at solar metallicity is modest and thus only weakly affects the evolution of the star. It is expected that the reduction in mass-loss rate due to magnetic mass-loss quenching will be modest. This is indeed the case, as shown by the non-rotating 15 M$_\odot$ models, the evolutionary tracks on the HRD are barely influenced by the magnetic mass-loss quenching (Figure \ref{fig:nrm1}, left panel); however, it should be noted that this does not mean that all stellar parameters are unaffected. For instance, small differences in stellar mass, age, and effective temperature are present throughout the main sequence (see Appendix Table \ref{tab:t1}). For example, the TAMS mass of model M1 (- - -) is 14.81 M$_\odot$, while for model M2 (- -B), it is 14.97  M$_\odot$. For comparison, in the case of the initially 80 M$_\odot$ non-rotating models computed by \cite{petit2017}, the TAMS masses of their non-magnetic and magnetic model with similar initial surface magnetic field strength are 52 M$_\odot$ and 62 M$_\odot$, respectively (see their Figure 7). Moreover, the magnetic model has a slightly shorter main sequence lifetime than the non-magnetic model (at the TAMS 11.20 Myr vs 11.26 Myr, respectively), which is also in agreement with the results of \cite{petit2017}.

In order to reliably compare these predictions to observations the following point should be considered: 
The rotational history of the star would need to be known to establish whether a non-rotating model may be applicable to compare with observations. If, for instance, the star had undergone significant magnetic braking in the past, then a non-rotating model may not be a reasonable choice for comparison with a currently slowly rotating star, since the previous rotational history may have changed the stellar properties. For example, \cite{meynet2011} pointed out that the core properties and the surface enrichment would be different. 

Since in a non-rotating case the assumption is that rotation-induced instabilities do not operate, surface nitrogen enrichment (middle panel, Figure \ref{fig:nrm1}) is only expected when the star reaches the red supergiant stage. Whether or not a surface magnetic field is present has no impact on these results in the present model so long as only mass-loss quenching is relevant. Nevertheless, the nitrogen abundance could help to observationally evaluate the applicability of non-rotating models.

The right panel of Figure \ref{fig:nrm1} shows the mass-loss history of models M1 (- - -) and M2 (- -B). Since mass-loss quenching reduces the effective mass-loss rate, the magnetic model will initially experience an order-of-magnitude lower mass-loss rate than the non-magnetic model. However, as the star evolves, so do the magnetic parameters (see Section \ref{sec:43}). As a consequence, the magnetic confinement weakens systematically as the star crosses the HRD. Ultimately, the magnetic field weakens to such a degree that mass-loss quenching becomes negligible. These results, although quantitatively dependent on the adopted mass-loss scheme (other applicable schemes for hot stars are derived by, e.g., \citealt{kudritzki1989,vink2001,muijres2012,krticka2014}), demonstrate the same qualitative behaviour, independent of the adopted mass-loss scheme. 

The initial reduction in the mass-loss rate of model M2 (- -B) results in a systematically higher stellar mass, hence a higher stellar luminosity is expected on the main sequence when compared to model M1 (- - -) due to mass-loss quenching (left panel of Figure \ref{fig:nrm1}, see also Figure 4 of \citealt{petit2017}). However, this expectation is limited, to a degree, by an interesting feedback effect. Since $\dot{M} \propto L_\star^x$, where $x$ is some positive power depending on the adopted scheme\footnote{For the scheme derived by \cite{dej1988} and used in this study, $x \sim 1.8$. For comparison, the \cite{vink2001} scheme yields $x \sim 2.2$.}, the mass-loss rate of the magnetic model will increase if it becomes more luminous. Therefore, one may conclude that mass-loss quenching alone results in the following loop of consequences:
\begin{enumerate}
\item Mass-loss quenching $\rightarrow$ initially lower mass-loss rate
\item Initially lower mass-loss rate $\rightarrow$ higher stellar mass
\item Higher stellar mass $\rightarrow$ higher stellar luminosity
\item Higher stellar luminosity $\rightarrow$ higher mass-loss rate
\end{enumerate}
In the case of the 15 M$_\odot$ models, mass-loss quenching alone is modest. However, \cite{petit2017} demonstrated that for higher-mass models, following the above reasoning, the consideration of surface fossil magnetic fields indeed leads to a notable increase in the stellar luminosity. Despite that this results in a consequent increase in the mass-loss rate of the magnetic model, the relative value of the mass-loss rate still remains lower compared to that of a non-magnetic model with the same initial mass. This is because the power that the luminosity depends on the stellar mass is higher than the power the mass-loss rate depends on the luminosity. 

%
%
%
\subsection{Non-magnetic vs magnetic models with rotation without \texorpdfstring{$D_{\rm ST}$}{DST}: models M3 (R- -) and M4  (R-B)}

Figure \ref{fig:nrm2} shows the predictions of models M3 (R- -) and M4 (R-B). Both of these models are initially rotating at $v_{\rm rot, ini} =$~200~km~s$^{-1}$ and are characterized by differential rotation. They do not include the viscosity of an ST dynamo for the internal angular momentum transport. Model M4 (R-B) includes mass-loss quenching and magnetic braking. Whether such a configuration, i.e. radial differential rotation, is possible for a star with large-scale surface fossil magnetic fields extending to the stellar interiors has not yet been established (however, see \citealt{duez2010}). It is likely that a strong magnetic field would flatten the $\Omega$ profile and result in near solid-body rotation, but perhaps not for the entire star. We further discuss this point in the last paragraph of Section \ref{sec:unc}.
 
The left panel of Figure \ref{fig:nrm2} shows models M3 (R- -) and M4 (R-B) on the HRD. It may seem intuitive that the above outlined mass-loss quenching loop yields a higher luminosity for model M4 (R-B) compared to model M3 (R- -). For reference, the TAMS masses of model M3 (R- -) and model M4 (R-B) are 14.71 M$_\odot$ and 14.94 M$_\odot$, respectively (see Appendix Table \ref{tab:t2}).

However, in this particular model setup the large increase of the stellar luminosity of model M4 (R-B) compared to model M3 (R- -) cannot be attributed to mass-loss quenching alone, even though the mass-loss rate of model M4 (R-B) with surface magnetic field is still lower than that of model M3 (R- -) until both models' effective temperatures reach below 10 kK (Figure \ref{fig:nrm2}, right panel). At 10 kK the surface magnetic field is so weak that mass-loss quenching becomes negligible.

Therefore, the feedback effects need a more careful interpretation in rotating models. We mention here that the rotating models do contain an additional scaling factor, which takes into account the rotational enhancement on the mass-loss rates, following the work of \cite{maeder2000}, namely Equation 4.29 in their work. As shown by other studies as well (e.g., recently \citealt{keszthelyi2017a}, their Figures 10 and 11), this enhancement factor on the mass-loss rates due to rotation is generally of the order of maximum a few percent, but becomes important when the surface rotational velocity approaches the critical velocity. Hence we checked whether this factor would play a role and found that in these model calculations the rotational velocities remain far from their critical values, ergo the rotational enhancement on the mass-loss rates (reaching a maximum of 3\%) is practically negligible in all models throughout their evolution.  

In contrast to the case of non-rotating models, rotating models that include surface magnetic fields account for magnetic braking. This mechanism leads to changes in the $\Omega$ profile near the stellar surface, which influences chemical mixing (Figure \ref{fig:nrm2}, middle panel) and hence it changes the mean molecular weight ($L_\star \propto\mu^x$, where $x$ is typically a high power, $x \sim 4$). When differential rotation is considered, magnetic braking induces significant chemical mixing via strong shears and therefore increases the average value of $\mu$, which in turn results in a higher luminosity. This can indeed be traced by the nitrogen enrichment of the model. As a consequence, the effects of magnetic braking alone could be described in the following manner for differential rotation: 
\begin{enumerate}
\item Magnetic braking $\rightarrow$ strong shear mixing
\item Strong shear mixing $\rightarrow$ increase of average $\mu$ and notable surface enrichment
\item Increase of average $\mu$ $\rightarrow$ higher stellar luminosity
\item Higher stellar luminosity $\rightarrow$ higher mass-loss rate
\end{enumerate}
Thus we confirm the findings of \cite{meynet2011} regarding the surface nitrogen enrichment. Although we changed the magnetic field evolution model ($B_p \propto R_\star^{-2}$, hence the magnetic field strength weakens with time), we observe the same qualitative behaviour: surface magnetic fields and radial differential rotation results in a very rapid enhancement of the surface nitrogen due to the large shears (Figure \ref{fig:nrm2}, middle panel). 

As a consequence of the induced mixing in the magnetic model, the ages of the models differ significantly. Even at the early phases, model M4 (R-B) takes a longer time to deplete core hydrogen and this difference is propagated to the TAMS age. The TAMS ages of models M3 (R- -) and M4 (R-B) are 13.45 Myr and 15.29 Myr, respectively (Appendix Table~\ref{tab:t2}). Therefore, while mass-loss quenching alone resulted in decreasing the age of the magnetic model, magnetic braking in differentially rotating models actually yields longer main sequence lifetimes.

To conclude, the combination of magnetic braking and differential rotation can explain the notably higher luminosity of model M4 (R-B) compared to M3 (R- -). Model M4 (R-B) becomes more luminous than model M3 (R- -) because the shears induced by magnetic braking increase the mean molecular weight on average inside the star. Additionally, mass-loss quenching also modestly increases the luminosity of model M4 (R-B) in comparison to model M3 (R- -). In fact, because magnetic braking efficiently leads to a higher luminosity of model M4 (R-B) compared to model M3 (R- -), the notably higher luminosity also scales the effective mass-loss rate of the stellar model. This is why the mass-loss histories of the two models (Figure \ref{fig:nrm2}, right panel) differ less than in a non-rotating case, even though mass-loss quenching reduces the effective mass-loss rate of model M4 (R-B).

%
%
%
\subsection{Non-magnetic vs magnetic models with rotation adopting \texorpdfstring{$D_{\rm ST}$}{DST}: model M5 (RS-) and M6 (RSB)}

Figure \ref{fig:nrm3} shows the predictions of models M5 (RS-) and M6~(RSB). Both models adopt an internal ST dynamo mechanism for the chemical element and angular momentum transport to ensure a flat angular velocity profile. Similar to model M4 (R-B), model M6 (RSB) includes mass-loss quenching and magnetic braking. 

The differences between these two models on the HRD are practically negligible, however other parameters show more noticeable differences. Throughout its evolution, model M6 (RSB) maintains a mass-loss rate which is significantly lower than that of model M5 (RS-), (Figure \ref{fig:nrm3}, right panel).

However, when comparing models M5 (RS-) and M6 (RSB), the nitrogen enrichment is notably lower in model M6 (RSB). In particular, the post-main sequence plateau remains 0.2 dex smaller compared to model M5 (RS-) when crossing the HRD (Figure \ref{fig:nrm3}, middle panel).
These results agree qualitatively\footnote{However, see Section \ref{sec:hunt} for the quantitative difference.} with the findings of \cite{meynet2011}, namely that when a flat angular velocity profile is assumed, and surface magnetic fields are considered, the mixing is much less efficient than in the case of differential rotation (cf. also model M4 (R-B) and model M6 (RSB) in the middle panels of Figure \ref{fig:nrm2} and \ref{fig:nrm3} with solid lines). Without magnetic braking, the solid-body rotating model results in higher surface nitrogen enrichment compared to a differentially rotating model (cf. also model M3 (R- -) and model M5 (R-B) in the middle panels of Figure \ref{fig:nrm2} and \ref{fig:nrm3} with dashed lines). When magnetic braking is considered, the surface angular velocity decreases rapidly, thus in solid-body rotating models the rotation of the entire star brakes. The systematically and uniformly decreasing value of angular velocity throughout the star leads to weaker chemical element transport (predominantly weaker meridional currents). This can be summarized in the case of solid-body rotation as follows: 
\begin{enumerate}
\item Magnetic braking $\rightarrow$ decrease of surface $\Omega$
\item Decrease of surface $\Omega$ $\rightarrow$ decrease of $\Omega$ throughout the star
\item Decrease of $\Omega$ throughout the star $\rightarrow$ weaker meridional currents
\item Weaker meridional currents $\rightarrow$ less chemical enrichment at the surface
\end{enumerate}
Hence magnetic braking results in lower meridional circulation velocity compared to a corresponding model without surface magnetic fields, and this is why the surface nitrogen enrichment of model M6 (RSB) is less than that of model M5 (RS-). 

Given these two differences between the models, that is, model M6 (RSB) is less enriched but evolves with a slightly higher mass than model M5 (RS-), they have an opposite effect on the stellar luminosity, thus the nearly identical evolutionary tracks on the HRD are explained. This is also the reason why the stellar ages are quite similar in this case, at the TAMS equal to 11.46 Myr and 11.36 Myr, respectively. For reference, the TAMS masses of model M5 (RS-) and model M6 (RSB) are 14.65 M$_\odot$ and 14.95 M$_\odot$ (see Appendix Table \ref{tab:t3}).

\begin{figure*}
\includegraphics[width=9cm]{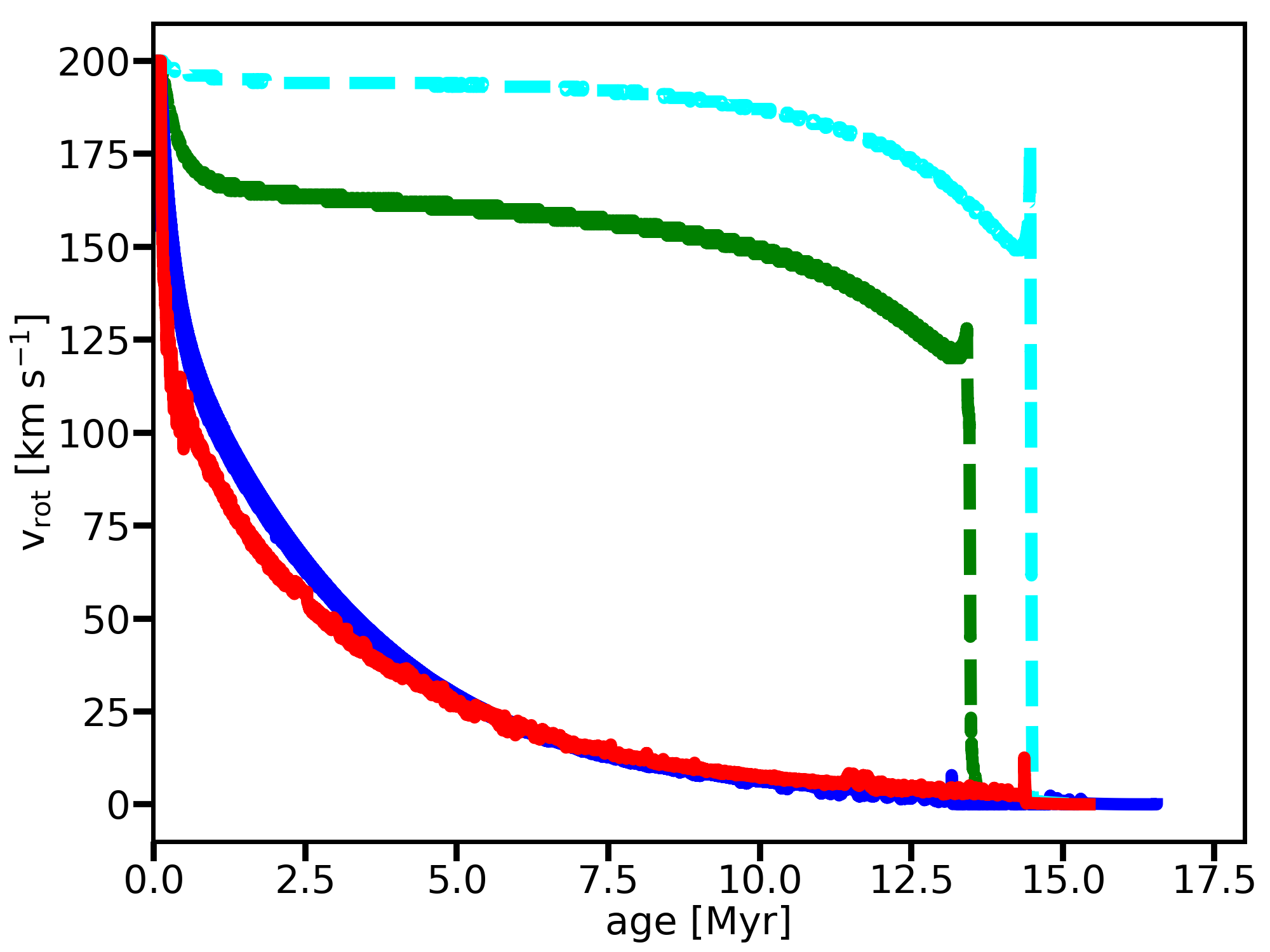}\includegraphics[width=9cm]{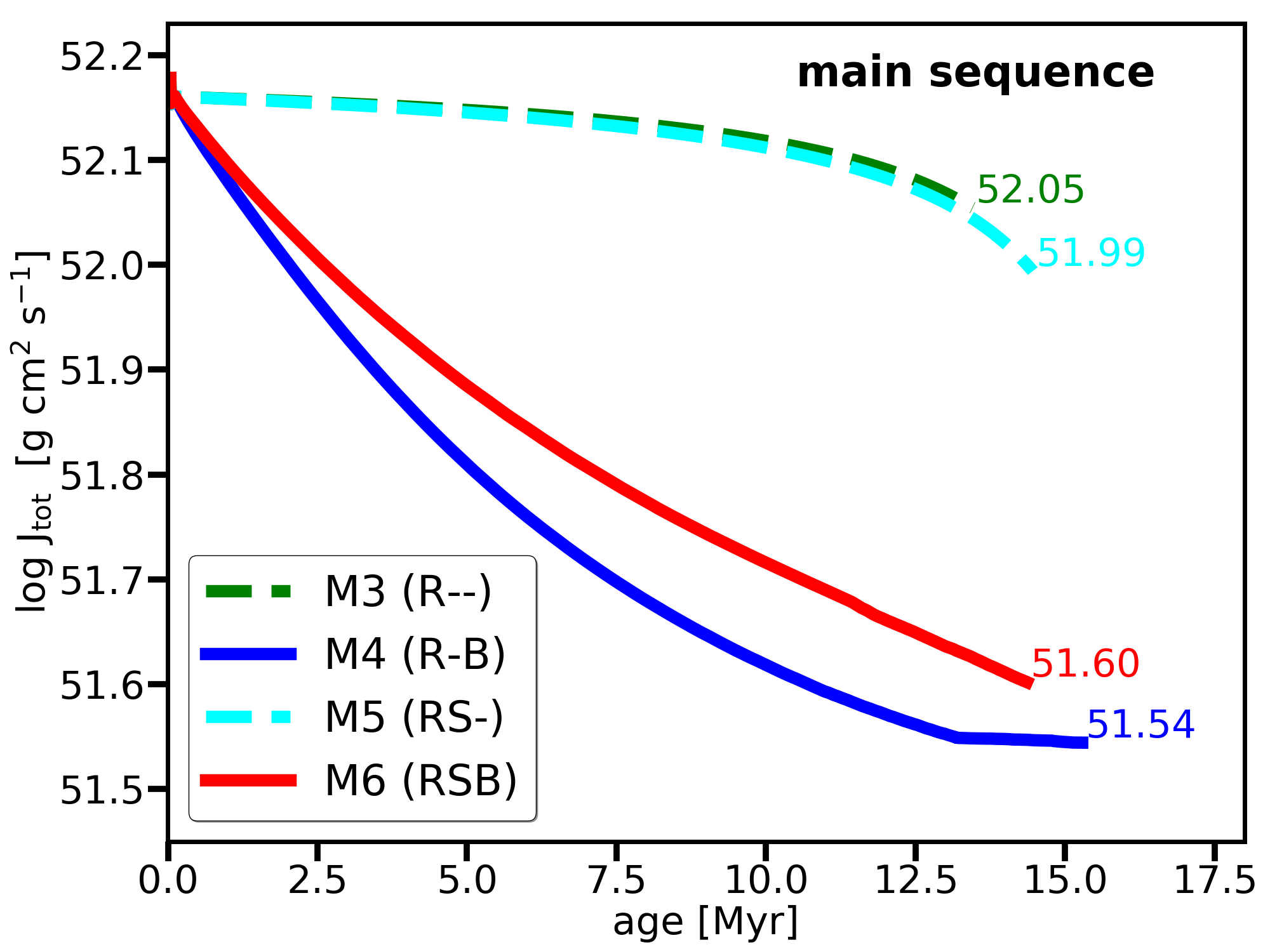} 
\caption{\textit{Left panel}: The time evolution of the surface equatorial rotational velocities from the ZAMS until the models start to deplete core Helium. \textit{Right panel}: The evolution of the total angular momentum over time on the main sequence. The TAMS values of log $J_{\rm tot}$ in units of g~cm$^2$ s$^{-1}$ are indicated next to the tracks.}\label{fig:rot1}
\end{figure*}
%
\section{Discussion}

\subsection{Rotational velocity and angular momentum evolution}

The left panel of Figure \ref{fig:rot1} shows the surface rotational evolution predicted by the 4 models with rotation and reveals that the rotational history of models with/without surface magnetic fields is very different. 

In the case of models M3~(R-~-) and M5 (RS-) (green and cyan lines), we observe the following: 
With the adopted initial mass and initial rotational velocity, differential rotation leads to a relatively modest decrease of the surface rotational velocity on the main sequence. However, this strongly depends on the initial mass and rotational velocity of the model, namely the surface rotation would brake much more rapidly if the model was more massive and/or had a higher initial rotational velocity \citep[e.g.,][]{georgy2013}.
When solid-body rotation is imposed by means of the viscosity of the ST dynamo, the rotational velocity remains nearly constant during (almost) the entire main sequence \citep{maeder2003,keszthelyi2017a}. The surface rotational velocity decreases abruptly after the TAMS since the radius of the star undergoes a dramatic increase.

In the case of models M4 (R-B) and M6 (RSB) (blue and red lines), we interpret their rotational history as follows:
When surface magnetic fields and hence magnetic braking is accounted for, the surface rotation of the star already brakes efficiently on the main sequence - as expected from theoretical works \citep{ud2009,meynet2011}. As a consequence, the surface rotational velocity rapidly approaches zero. 

However, the time-scale of magnetic braking depends on the assumption of how angular momentum is distributed inside the star, and how efficiently the surface angular momentum reservoir can be replenished by driving angular momentum from the core. The problem of internal angular momentum distribution was already addressed by \cite{pin1989}, who concluded that it may significantly influence the evolution of the surface angular momentum reservoir. 
Model M6 (RSB) undergoes a somewhat more rapid initial spin-down than model M4 (R-B) although, despite the initially more rapid spin-down, both curves are qualitatively similar. The time-scale required to slow down the surface rotation from the ZAMS (v$_{\rm rot, \rm ini}$  = 200 km~s$^{-1}$) to reach a surface rotational velocity below 50~km s$^{-1}$ is 3.3~Myr and 2.8 Myr for models M4 (R-B) and M6 (RSB), respectively. We will call this the `effective spin-down time', meaning the time after ZAMS to become a slow rotator ($< $50 km s$^{-1}$), in order to identify the time after which surface rotation has small or negligible impact on the structure and evolution of the model. 

The right panel of Figure \ref{fig:rot1} shows the evolution of the total angular momentum of the star models on the main sequence. 
Models M3 (R- -) and M5 (RS-) lose a small amount of total angular momentum on the main sequence. This loss is due to mass loss, which reduces the angular momentum content of the star \citep{langer1998, vink2010,keszthelyi2017a}. 
When surface magnetic fields are accounted for, they strongly impact the angular momentum evolution of the star since magnetic braking is a very efficient mechanism to remove angular momentum \citep[see also][]{song2018}. Despite the fact that while on the main sequence model M6 (RSB) has an order of magnitude lower mass-loss rate than model M5 (RS-)  (see Figure \ref{fig:nrm3}, right panel), the former model loses almost 0.4 dex more of its total angular momentum. This is because magnetic braking is significantly more efficient to remove angular momentum from the star than mass loss by the stellar winds \citep{ud2009}. 

It may not be immediately intuitive why model M4 (R-B) loses more of its total angular momentum on the main sequence than model M6 (RSB), in particular because one may expect that solid-body rotation brakes the rotation of the entire star (i.e., reduces the total angular momentum reservoir), while differential rotation only allows for effectively braking the surface (i.e., exhausting only the surface angular momentum reservoir). Hence, if the only difference between the two models was their rotational configuration, model M4 (R-B) would be expected to maintain more of its total angular momentum. However, this is not the case: model M4 (R-B) does lose more angular momentum than does model M6 (RSB). This is because model M4 (R-B) evolves with a higher luminosity than all the other models due to its increased mean molecular weight. As a consequence, its mass-loss rate is higher compared to model M6 (RSB). Therefore, while magnetic braking is indeed the main driver of angular momentum loss, model M6 (RSB) loses $\sim$ 0.06 dex less of its total angular momentum at the TAMS compared to model M4 (R-B) because its mass-loss rate is lower. 

\begin{figure*}
\includegraphics[width=19cm]{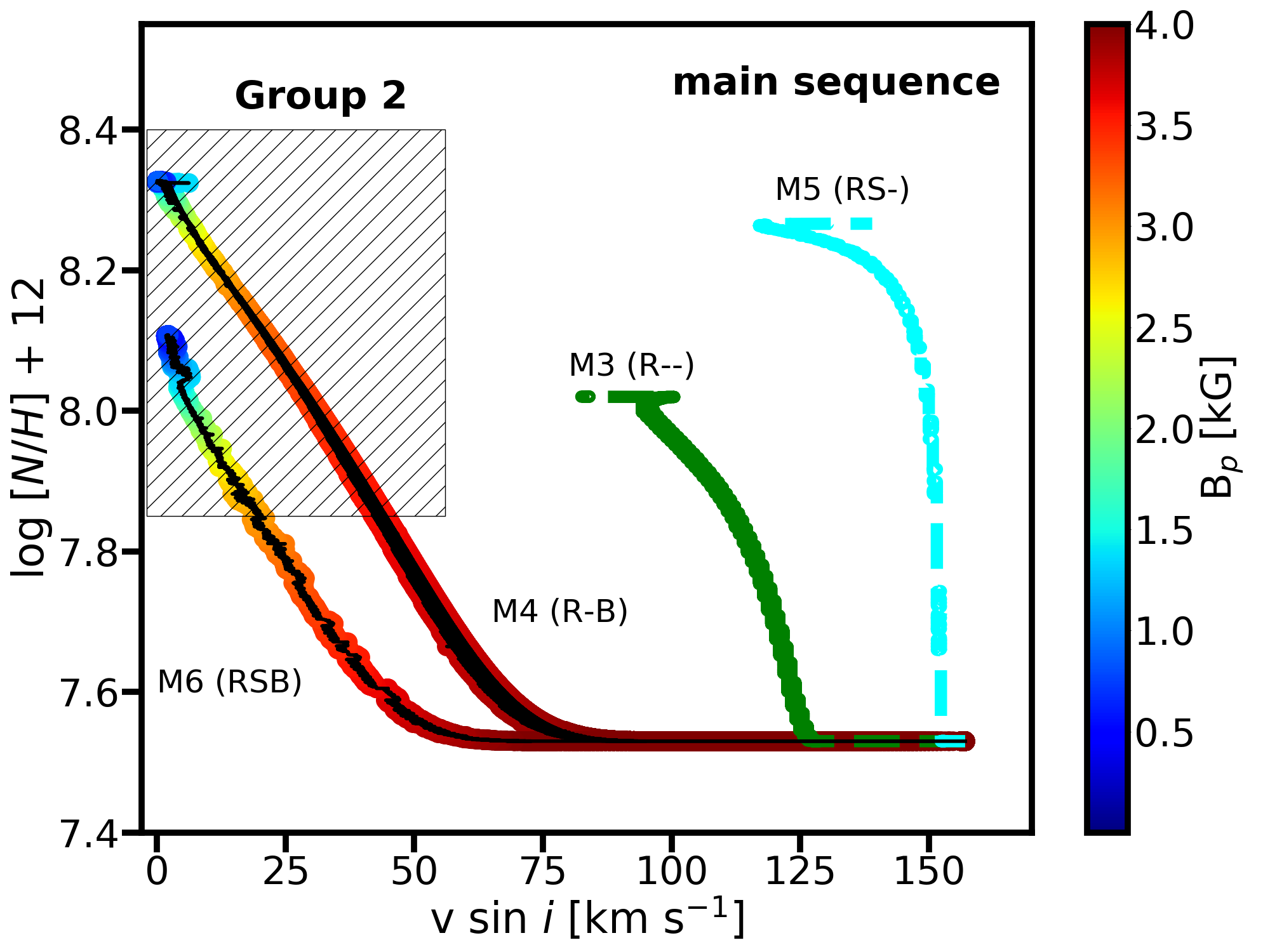}
\caption{Hunter diagram of the main sequence phase of the rotating models. $v$ sin $i$ is obtained by scaling the surface rotational velocities by $\pi / 4$. Group 2 stars are denoted with the hatched box on the diagram and the polar magnetic field strength is colour-coded in the case of the two magnetic models. The evolution of the models begins at the lower right corner of the diagram.}\label{fig:hun}
\end{figure*}

\subsection{Surface nitrogen enrichment}\label{sec:hunt}

The surface nitrogen abundance and the projected rotational velocity (plotted on the Hunter diagram) are two very practical quantities to trace and to study rotational mixing in massive stars since rotational mixing brings core-processed CNO elements to the stellar surface \citep{maeder2014a}. While in Section 3 we focused on the predictions of the models and the inferred impact of a surface magnetic field, now we turn our attention to the relevance of surface magnetic fields in the context of the Hunter diagram \citep{hunter2008}.

The VLT-FLAMES Surveys \citep{evans2004,evans2005,evans2008} obtained a significant collection of spectroscopic data of hot, massive stars in the Large Magellanic Cloud (LMC), Small Magellanic Cloud (SMC), and our Galaxy. From these observations several studies have identified anomalies in the Hunter diagram that cannot be explained with `standard rotational mixing' \citep{hunter2008,hunter2009,brott2009,brott2011b,maeder2014a,grin2017,dufton2018}.

In particular, two groups of observed stars (see Figure 1 of \citealt{hunter2009}) are challenging to explain with standard stellar evolution models (see also \citealt{aerts2014}). Group 1 stars exhibit fast rotation but show negligible evidence of surface chemical enrichment. Group 2 stars are inferred to be slow rotators but have notable surface enrichment. The stars in both groups have surface gravities that are generally consistent with them being main sequence stars. The existence of both groups is somewhat counterintuitive since typically the faster the star rotates, the more efficient is the chemical mixing, thus the larger is the surface chemical enrichment. Therefore, standard stellar evolution models, such as models M3 (R- -) and M5 (RS-), do not seem to be able to explain such a behaviour in main sequence stars. 

While comprehensive observational data supports the existence of these groups in the LMC \citep[e.g.,][]{hunter2008,brott2011b,rivero2012a,rivero2012b,grin2017,dufton2018}, the same anomalous behaviour, notwithstanding statistically smaller samples, is also evidenced in our own Galaxy \citep{crowther2006,morel2008,hunter2009,przybilla2010,briquet2012,aerts2014,martins2015,cazorla2017a,markova2018}.

Surface magnetic fields have been invoked to account for the phenomenon \citep{morel2008,hunter2009,brott2011b,martins2012,aerts2014,martins2015,grin2017,dufton2018} and, in particular, the mechanism of magnetic braking has been proposed to explain Group 2 stars \citep{meynet2011,potter2012b}, although binarity may also provide an alternative explanation \citep[e.g., most recently,][]{song2018b}.

Here we aim to focus only on Galactic Group 2 stars, and the general behaviour of the computed models as shown in Figure \ref{fig:hun}. In contrast to models M3 (R- -) and M5 (RS-), both models M4 (R-B) and M6 (RSB) exhibit rapid decrease of their surface rotation on the main sequence. We recall that the effective spin-down time for these models to become slow rotators is of the order of 3 Myr. For model M6 (RSB) the time after the ZAMS that is required to start increasing the surface nitrogen abundance is 2.0 Myr (when $v \sin i \sim 50$~km~s$^{-1}$). In the case of model M4 (R-B) the enrichment takes place on a shorter time-scale, the surface nitrogen abundance increases after 1.1 Myr (when $v \sin i \sim 80$ km s$^{-1}$). As is common, the surface equatorial rotational velocity is converted to projected rotational velocity by accounting for an averaged inclination, i.e., $v_{\rm rot} \cdot \pi /4 \sim v \sin i$. 

Not only does model M4 (R-B) mix more rapidly than model M6 (RSB), but the overall surface nitrogen abundance is higher too (see also middle panels of Figure \ref{fig:nrm2} and Figure~\ref{fig:nrm3}). With [N/H] we denote the number fraction of nitrogen relative to hydrogen. At the TAMS, log [N/H] + 12 = 8.33 in the case of model M4 (R-B) and log [N/H] + 12 = 8.11 in the case of model M6 (RSB), thus there is an overall 0.2 dex difference which originates from the assumptions regarding the internal rotation profile of the star. 

Interestingly, this difference is also observable in models without surface magnetic fields, however, in the opposite direction. Model M5 (RS-) reaches log~[N/H]~+~12~=~8.27 at the TAMS, while model M3 (R-~-) produces log~[N/H]~+~12~=~8.02. This is because when surface magnetic fields are not considered, $\Omega$ is maintained near to uniform throughout the star model M5 (RS-). As a consequence, meridional currents remain very efficient in transporting core-processed chemical elements to the stellar surface.

Therefore, comparing models with different rotational configuration, we can conclude the following: magnetic braking greatly enhances mixing if radial differential rotation is allowed for in the model. On the other hand, the inclusion of surface magnetic fields yields a lower enrichment in the case of near solid-body rotation. 

This result reinforces the point that although magnetic braking does provide a qualitative explanation for Group~2 stars, the surface nitrogen enrichment depends strongly on the internal mixing processes. This could be the reason why previous studies by \cite{morel2008}, \cite{aerts2014} and \cite{martins2015} did not find a direct correlation between magnetic field strength and measured nitrogen abundance. This might imply that magnetic massive stars may have different angular momentum transport processes at work or the strength of the transport varies significantly from one star to another.
 
Our results, for the case of differential rotation, are in complete agreement with findings of \cite{meynet2011} regardless of the additionally considered magnetic field evolution and mass-loss quenching in this study. However, \cite{meynet2011} obtained a different conclusion considering magnetic braking and solid-body rotation jointly. Namely, without mass-loss quenching and magnetic field evolution, their solid-body rotating models undergoing magnetic braking did not produce notable surface nitrogen enrichment, and thus did not appear as Group 2 stars on the Hunter diagram. In this study, however, model M6 (RSB) does yield an observable surface nitrogen enrichment. This may have partially been a result of the slight difference in the adopted initial mass of the models. \cite{meynet2011} considered $M_{\star, \rm ini} = 10$ M$_{\odot}$, while in this work we considered $M_{\star, \rm ini} = 15$ M$_{\odot}$. Therefore, we computed a model with the exact same configuration as model M6 (RSB) but with $M_{\star, \rm ini} = 10$ M$_{\odot}$ and found that it overlaps with model M6 (RSB) on the Hunter diagram. Thus, this difference is only due to considering mass-loss quenching and magnetic field evolution which lead to weakening magnetic braking over time.
 
Since Group 2 stars might provide indirect evidence for the presence of magnetic stars in extragalactic environments, further spectropolarimetric measurements of \textit{Galactic} Group 2 stars would be valuable to establish the robustness of this potential proxy. Therefore, an important question remains the incidence rate of surface magnetism in Galactic Group~2 stars. In particular, stars with known nitrogen enrichment and relatively modest rotational velocities, e.g., from the sample of Galactic O stars by \cite{markova2018}, and from the samples of Galactic B stars by \cite{crowther2006} and \cite{fraser2010} should be investigated. We also mention that the modest rotational velocities are an advantage to search for surface magnetic fields through spectropolarimetric observations since in those cases the spectral line broadening due to rotation requires a shorter exposure time to detect a magnetic field of given strength.

Since we accounted for the evolution of the surface magnetic field (colour-coded in Figure \ref{fig:hun}), we can place some constraints regarding this matter. Both models with surface magnetic fields are assumed to have an initial polar magnetic field strength of 4 kG. During the early evolution, this field should remain strong if magnetic flux conservation is valid. The field starts weakening once the stellar radius increases. Since typically the change in stellar radius is modest on the main sequence (a factor of 2 - 3), one would expect that the magnetic field strength also weakens modestly (a factor of 4 - 9). Indeed, both models M4 (R-B) and M6 (RSB) approach Group 2 on the Hunter diagram with $B_{\rm p} \sim$ 3.5 and 3.0~kG, respectively, and at the TAMS their magnetic field strengths are $B_{\rm p} \sim$ 870 G and 740 G, respectively. Since the adopted ZAMS field strength is typical of known magnetic O and B stars, these values provide some guidance as the expected observable field strength in Group 2 stars. 
 
The spectropolarimetric measurements must be sufficiently sensitive to detect a polar field strength of $B_{\rm p} \sim$ 800~G (if the star is close to the TAMS), that is, a peak longitudinal field of $B_z \sim$~250~G, assuming that the choice of $B_{\rm p,ini} =$ 4 kG at the ZAMS and the model of magnetic flux conservation are appropriate. Detecting such a field at $3\sigma$ would require a longitudinal field uncertainty (i.e., 1$\sigma$ error bar) of about 80~G. Weaker initial magnetic fields would naturally imply better required precision (e.g., 40~G for an initial polar field of 2~kG). These values are comparable with those obtained for the O and B samples of the MiMeS survey (respectively 50~G and 30~G; \citealt{grunhut2017} and Wade et al., in prep.). 

In addition to the obtaining spectropolarimetric measurements of known Galactic Group 2 stars, it may be worthwhile to identify stars in the MiMeS survey located in Group~2 of the Hunter diagram (e.g., exploiting abundances and $v\sin i$ such as those reported by \citealt{martins2015} for the MiMeS O-type stars) and evaluate if the precision with which they were individually observed is sufficient to detect their expected weaker magnetic fields (and re-observe them with better precision if necessary). 

Ultimately, magnetic fields and binarity are not necessarily mutually exclusive to explain Group 2 stars on the Hunter diagram. While other studies (e.g., \citealt{song2018b}) have shown that binarity can explain Group 2 stars, we have shown here that surface magnetism in single stars can also explain them.
The question that should be answered by observations is: \textit{What fraction of those stars are in binary systems and what fraction of them possess detectable surface magnetic fields?}

\begin{figure*}
\includegraphics[width=9cm]{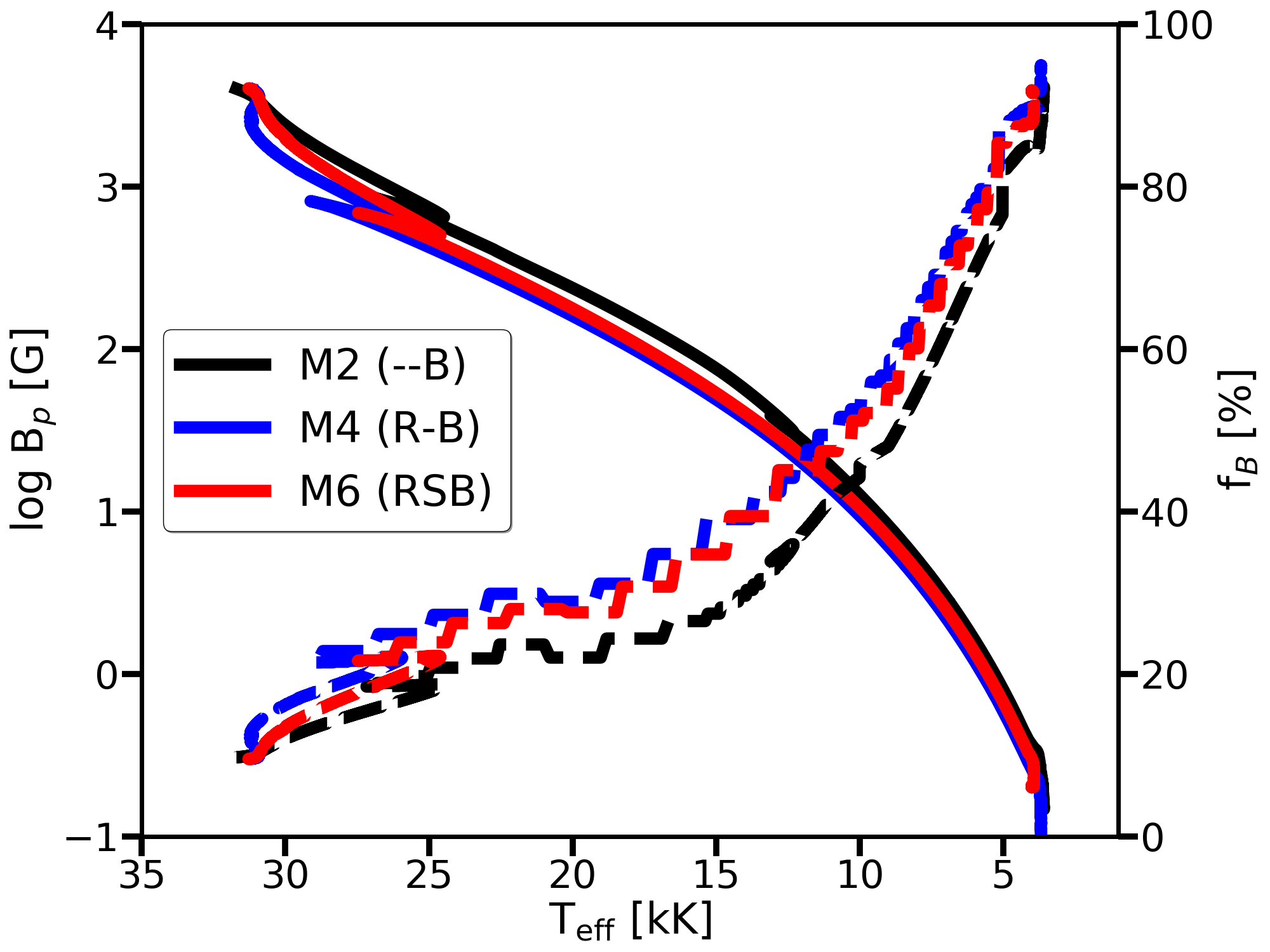}\includegraphics[width=9cm]{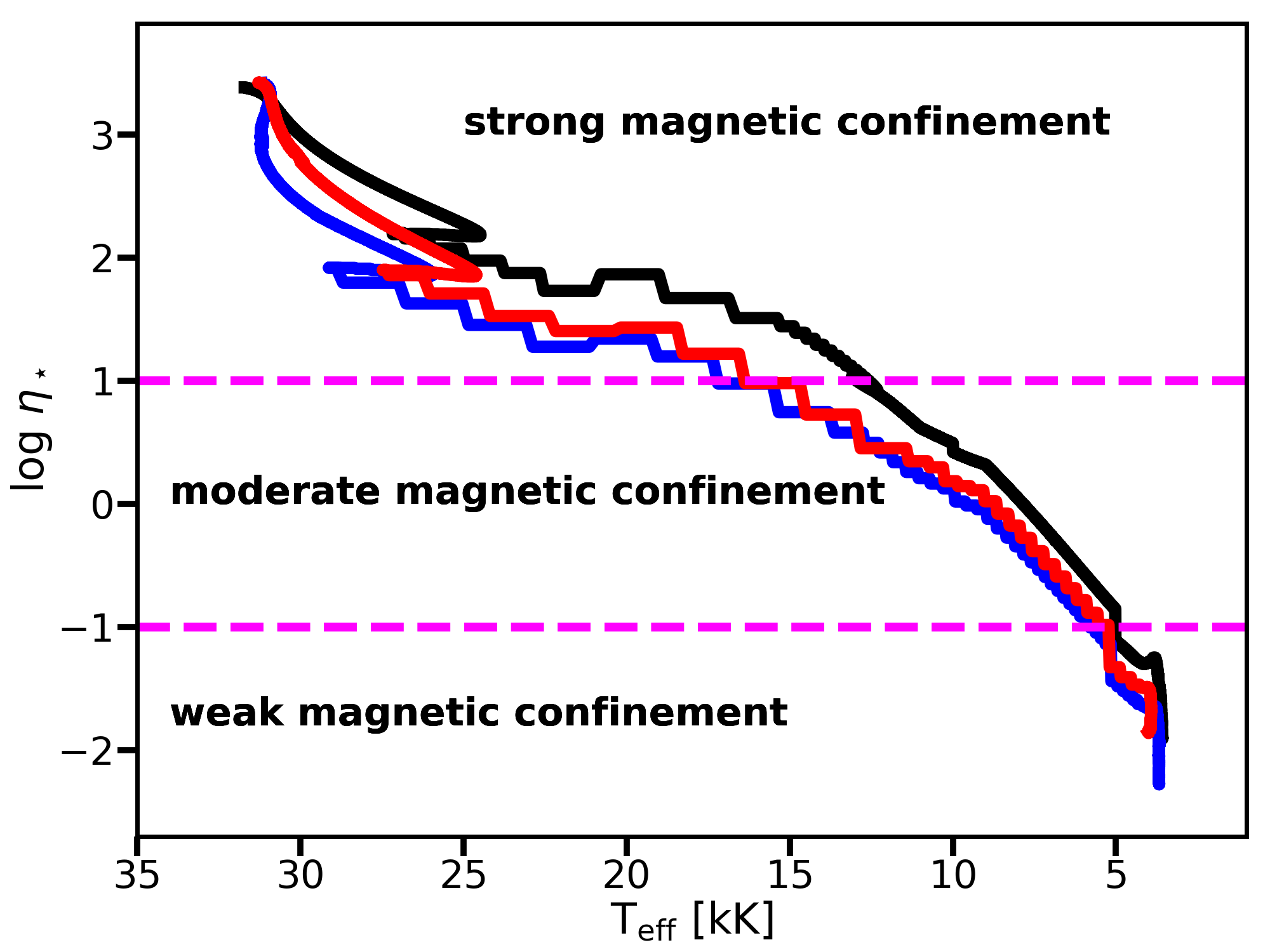} 
\caption{Evolution of the surface magnetic parameters. \textit{Left panel:} log of the polar magnetic field strength (left abscissa, solid lines) and the escaping wind fraction (right abscissa, dashed lines) against the effective temperature. \textit{Right panel:} evolution of the equatorial magnetic confinement parameter. }\label{fig:mag}
\end{figure*}

\subsection{Evolution of the surface magnetic field strength and magnetospheric parameters}\label{sec:43}

As both the magnetic field strength and mass-loss rate evolve with time, the nature of the interaction between the surface magnetic field and the stellar wind is expected to change over evolutionary time-scales. 

Figure \ref{fig:mag} shows the evolution of the magnetospheric parameters, $B_p$, $f_B$, and $\eta_\star$ (see Equations \ref{eq:quench} and \ref{eq:eta}). The left panel reveals that the main sequence weakening of the polar magnetic field strength (left abscissa and solid lines) is relatively modest, under the assumption that the magnetic flux is conserved during the evolution. However, once the stellar radius increases drastically after the TAMS, the magnetic field strength rapidly weakens from the order of kG to hundreds of G at $\sim$ 25 - 15 kK, to the order of 100 G to 10 G between 15 and 10 kK, and to the order of 10 G to 1 G between 10 and 5 kK. Since the stellar radius governs the field evolution, models M2 (- -B), M4 (R-B), and M6 (RSB) show similar characteristics because the evolution of the stellar radius is similar in these models. 

The left panel of Figure \ref{fig:mag} also shows the escaping wind fraction $f_B$ (right abscissa and dashed lines). On the main sequence, the escaping wind fraction is 10 - 20\%, which means a drastic reduction in the mass-loss rates for stars with surface magnetic fields. Until the late B star regime ($\sim$ 10 kK) mass-loss quenching is still very efficient as $f_B \approx$ 20 - 50\%. Therefore during the O-star regime in our models the returning wind fraction (that is, 1~-~$f_B$) is of the order of 80 - 90\%, which means that most of the line-driven mass loss returns to the stellar surface due to channeling by the magnetic field through closed field lines, while for B stars this fraction is at least 50\%. Evidently, this ratio depends on the assumption of the initial magnetic field strength. In extreme cases, the returning mass fraction can go up to 96\% as in the case of NGC 1624-2 (the escaping wind fraction is $f_B = 4\%$, \citealt{petit2017}), the most strongly magnetized known O-type star, which is believed to host a giant magnetosphere \citep{wade2012,petit2015,erba2017,daviduraz2018}.

The right panel of Figure \ref{fig:mag} shows how the equatorial magnetic confinement parameter evolves over time as a function of effective temperature. This means that during the early evolution of the star the magnetic confinement is expected to be strong ($\eta_\star > 10$) for a typical initial surface magnetic field strength. As shown previously by \cite{keszthelyi2017b}, $\eta_\star$ can be very sensitive to changes in $\dot{M}$ on the main sequence. This is especially the case when sudden changes in the mass-loss rates are considered due to the bi-stability mechanism \citep{pauldrach1990,vink1999,petrov2016,keszthelyi2017a,sander2018,vink2018}.

However, if the mass-loss rates are not assumed to show abrupt changes (see the right panels of Figures \ref{fig:nrm1}-\ref{fig:nrm3}), then the evolution of the magnetic confinement parameter is characterized by a systematic decrease in $B_p$ along with a systematic increase in $\dot{M}$. According to our calculations of a typical massive star model, at $T_{\rm eff} \sim 20 - 8 \, \mathrm{kK}$ a moderate magnetic confinement ($\eta_\star \approx 0.1 - 10$) is still retained. Ultimately, the stellar wind dominates over the magnetic field, which weakens as the star becomes a red supergiant, and as a consequence, the initially strong magnetic confinement ($\eta_\star \approx 10^3$) disappears in these models when they cross the HRD.

Most observed magnetic hot stars are believed to be on the main sequence, however these results can be strongly influenced by the particular evolutionary models chosen for comparison, especially since most determinations rely on non-magnetic stellar evolution models. The actual position of the TAMS for these stars thus remains to be determined.  
In stellar evolution models, along with the efficiency of rotationally-induced instabilities and thus chemical mixing in the envelope, a key parameter that influences the position of the TAMS on the HRD is the value of core overshooting \citep{bressan1981,stothers1985,langer1986,maeder1991,higgins2019}. 
For instance, \cite{castro2014} remark that, in general, core overshooting may require a different calibration depending on the initial mass of the star, while grids of stellar evolution models typically use one calibrated value for an entire mass range (\citealt{schaller1992,brott2011};\break \citealt{ekstroem2012}, but see also \citealt{vandenberg2006}). Moreover, \cite{briquet2012} argued that fossil fields could suppress core overshooting, and \cite{petermann2015} included an \textit{ad hoc} reduction in the convective core size to explain observed properties of B supergiants. The choice of overshoot parameter determines the width of the main sequence, which in turn may change the derived ages and evolutionary status of observed magnetic stars. 
 

To observationally constrain the fossil field evolution scenario, further spectropolarimetric measurements are required of stars that are presumably evolved. Non-chemically peculiar B and A supergiants (in the range of T$_{\rm eff} \sim~25 - 8$~kK) are of interest in this respect, since, if magnetic flux conservation is valid, a fraction of those stars should have observable fields consistent with their progenitors in the O and early B phase. Indeed, a few of these objects have already been discovered, however examples remain scarce. 
 
\cite{fossati2015b} report the detection of surface magnetic fields on the order of 10-100 G in the early B stars $\beta$~CMa (B1II, $T_{\rm eff} = 24.7$~kK, $M_\star = 12$ M$_\odot$, $B_p = 100$~G,) and $\epsilon$~CMa (B1.5II, $T_{\rm eff} = 22.5$~kK, $M_\star = 13$ M$_\odot$, $B_p > 13$~G). Three non-chemically peculiar A-type supergiants are known to possess weak surface magnetic fields, namely $\iota$ Car (A7Ib, $T_{\rm eff} = 7.5$~kK, $M_\star = 6.9 - 11.0$ M$_\odot$, $B_p = 3$~G, \citealt{neiner2017}), HR 3890 (A7Ib, $T_{\rm eff} = 7.5$~kK, $M_\star = 10.9-14.7$ M$_\odot$, $B_p = 6$ G, \citealt{neiner2017}), and 19 Aur (A5Ib, T$_{\rm eff} = 8.5$~kK, $M_\star = 6.9 - 9.7$ M$_\odot$, B$_p = 3$~G, \citealt{martin2018}). It is currently unclear whether these fields of A supergiants are organized fossil fields or generated by a near-surface dynamo mechanism. These stars might belong to the group that exhibit Vega-like magnetism as detected in intermediate-mass stars \citep{lignieres2009}. Since the mass determination of these objects is highly uncertain (mostly because their rotational histories are not known and partially because non-magnetic stellar evolution models were used for comparison), it is possible that these are intermediate-mass stars. Nevertheless, a class of evolved hot massive stars with weak surface magnetic fields should be observable, quite similar to these detections, but even in the $>$~15~M$_\odot$ initial mass range.

\subsection{Magneto-rotational evolution}\label{sec:mre}
%
\begin{figure*}
\includegraphics[width=19cm]{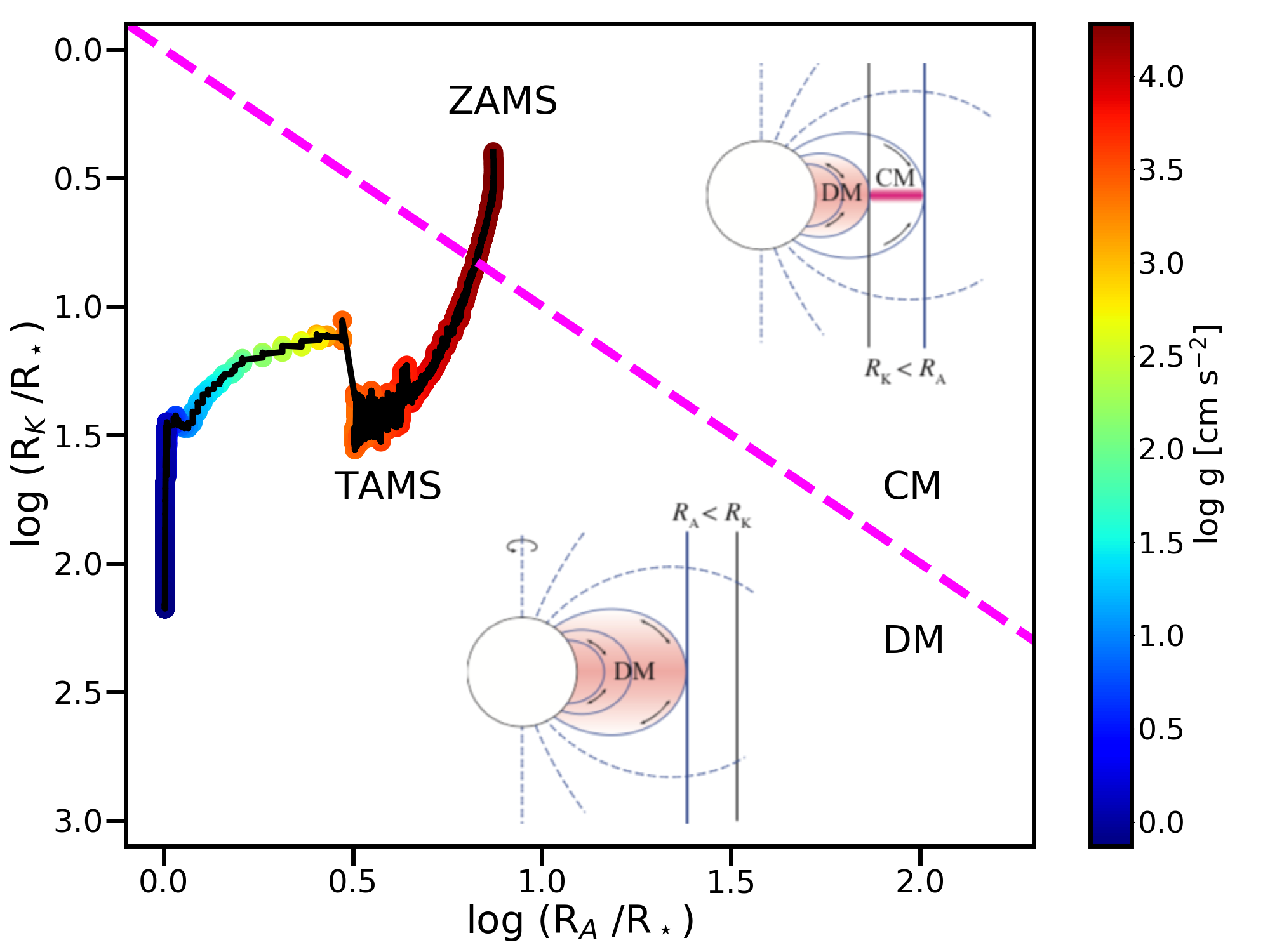}
\caption{Evolution of a magnetic model M6 (RSB) on the R$_K$-R$_A$ plane, colour-coded with log g. As the star spins down, it crosses the dashed magenta line which represents the transition from the CM regime to the DM regime. The Alfv\'en radius decreases rapidly after the TAMS. The two cartoons are adopted from \citet{petit2013}.}\label{fig:mag2}
\end{figure*}

Magnetospheres of hot massive stars are classified by \cite{petit2013} into dynamical magnetospheres (DM) and centrifugal magnetospheres (CM), depending on the relative sizes of the Alfv\'en radius and Kepler co-rotation radius. The Alfv\'en radius $R_A$ is approximated by \cite{ud2002} as 
\begin{equation}
\frac{R_A}{R_\star} = 0.29 + (\eta_\star + 0.25)^{0.25} \, ,
\end{equation}
and the Kepler co-rotation radius is expressed as
\begin{equation}
\frac{R_K}{R_\star} = \left( \frac{v_{\rm rot}}{\sqrt{G M_\star / R_\star}} \right)^{-2/3} \, .  
\end{equation} 
A magnetosphere is classified as a DM when $R_A < R_K$, and as a CM when $R_K < R_A$. 
A DM contains wind material, driven from the surface and confined by closed magnetic loops, which co-rotate with the stellar surface, and then shocks near the magnetic equator before falling back onto the star, creating dynamically complex flows (e.g., \citealt{babel1997,ud2002,owocki2016}). In the case of rapidly-rotating stars, rotation plays a significant dynamical role as the additional centrifugal support past the Kepler co-rotation radius can lead to the accumulation of wind material trapped at the apex of the closed loops, forming a CM \citep{townsend2005}. Rotationally-modulated variations in a number of observational diagnostics (Balmer lines, UV, X-rays, etc.) have indeed been linked to the presence of such magnetospheres (e.g., \citealt{landstreet1978,shore1990,gagne2005,petit2013}).

Figure \ref{fig:mag2} shows the evolution of model M6 (RSB) on the $R_K$-$R_A$ plane, the confinement-rotation diagram (see also Figure 3 of \citealt{petit2013}). Model M4 (R-B) follows a nearly identical evolution on this diagram with a somewhat larger noise close to the TAMS, thus for clarity only one model is shown. The star begins its main sequence evolution in the CM regime given the assumed initial surface magnetic field strength and rotational velocity. However, as the surface rotational velocity decreases, the Kepler radius increases.  In the case of model M6 (RSB) the transition from the centrifugally supported magnetosphere regime to the dynamical magnetosphere regime occurs after the star becomes a slow rotator ($v_{\rm rot} <$ 50 km~s$^{-1}$) at 3.4 Myr after the ZAMS. At that time the Alfv\'en and Kepler radii are both 6.7 R$_\star$ (or roughly 34~R$_\odot$). It is presently unknown how this transition from the CM to DM regime could be properly characterized, and whether the wind material trapped in the centrifugally supported region would mostly fall back onto the star, would escape, or perhaps may be forming a disk as the transition occurs. New magnetohydrodynamic simulations will study the transition in more detail (ud-Doula, priv. comm.). 

The star then spends the remainder of its evolution in the DM regime. An abrupt decrease in $R_K$ occurs after the TAMS, due to the sudden increase of the stellar radius. Nevertheless, $R_K$ systematically increases with time, while $R_A$ systematically decreases. The former is mainly governed by the evolution of the rotational velocity on the main sequence as long as changes in $M_\star$ and $R_\star$ are modest. After the TAMS, the increase of $R_K$ is attributed to the decreasing stellar mass. The evolution of $R_A$ depends on the evolution of the magnetic confinement parameter $\eta_\star$ (see Figure \ref{fig:mag}). There is a stable decrease on the main sequence from $R_A =$ 7.7 R$_\star$ at the ZAMS to $R_A =$ 3.3 R$_\star$ at the TAMS. After the star crosses the HRD, the initially large-scale magnetosphere has no or minimal effect on the stellar atmosphere. A key take-away of the CM to DM evolution is that (although quantitatively depending on the initial mass, rotation rate, field strength, etc.) by identifying a centrifugally-supported magnetosphere, constraints can, in principle, be placed on the age of the star. Since the surface spin-down of more massive stars is more rapid, magnetic O stars in the CM regime must be very young, close to their ZAMS\footnote{Or alternatively those stars may have been spun-up by, e.g., mass transfer in a binary system; see \citealt{grunhut2013}.}. On the other hand, in case of observed DM stars with $R_K$ / $R_A$ $>$ 5, there may be a strong indication that the star is evolved and is possibly on the post-main sequence. This is consistent with the fact that those stars that fulfill this criterion typically have lower values of observed log $g$ than those that have smaller $R_K$ / $R_A$ ratios \citep[see Table 1 and Figure 3 of][]{petit2013}. Additionally, we mention that this method may be worth comparing with cluster ages when available (see also \citealt{schneider2016}). It may also prove to be a robust tool to estimate the age and evolutionary status of runaway magnetic stars, e.g., HD 57682 (see \citealt{comeron1998} and \citealt{grunhut2009}).

\subsection{Uncertainties of fossil field evolution and limitations of the current models}\label{sec:unc}

The main sequence is the longest and most stable amongst all burning stages in a star's lifetime. In single stars, it is expected that fossil fields would remain stable, but their strength generally weakens on the main sequence. Observations show evidence that these large-scale fossil fields are stable during decades \citep[e.g.,][]{oksala2012,silvester2014,shultz2016,sikora2018}.

The adopted model of surface magnetic field evolution follows empirical constraints: the fields are strong on the main sequence ($B_p \approx 10^2 - 10^3 \, \mathrm{G}$), but they show evidence of weakening as stars age \citep{landstreet2007,landsreet2008,blazere2015,grunhut2017}. However, it remains debated whether the model of magnetic flux conservation is valid or not. Theoretical works \citep[e.g.,][]{braithwaite2017} point out that Ohmic decay is expected to lead to a more rapid decrease of B$_p$ than would result from magnetic flux conservation. However, the magnetic flux of compact remnants, in particular, of neutron stars is consistent with a constant magnetic flux evolution of their progenitor OB stars \citep{ferrario2008}. 
There are observational studies \citep{landstreet2007,fossati2016} that infer a field evolution model consistent with magnetic flux decay, on the other hand, other observational studies argue for the validity of magnetic flux conservation \citep{grunhut2017,neiner2017}. Recently \cite{shultz2018} proposed that the magnetic flux may depend on the stellar mass. To a large extent, this debate is a consequence of the lack of comprehensive observational data of evolved massive stars with sufficient precision to reliably test the different models (Petit et al., in prep.).

During the post-main sequence phases two key phenomena occur that should determine the fate of a surface fossil field: i) development of convective zones in the stellar envelope, and ii) the rapid change in the stellar radius. 

A major complexity is to consider the interplay between convective zones and a fossil magnetic field. Since this interaction is not well understood, we did not include the suppression of developing convective zones. This phenomenon is expected for strong magnetic fields, however during the post-MS phase, the weaker fossil fields may not suppress these zones. Additionally, convective regions can give rise to dynamo activity, generating small-scale magnetic fields, which can interact with the fossil field (see, e.g. \citealt{feat2009}). The current challenges can be listed as follows: 
\begin{itemize}
\item Parametric, one-dimensional prescriptions are not established to account for how a fossil field and convection would interact in massive stars.
\item One-dimensional hydrodynamical codes treat convective layers in a simplified approach. Usually, the mixing-length theory and solid-body rotation are adopted in convective zones. 
\item Convection-driven, dynamo-generated magnetic fields are not yet ubiquitously implemented in stellar evolution codes. However, recent studies of the cores of evolved massive stars 
(e.g., \citealt{maeder2014b,cantiello2016,kissin2018}) have begun to explore such effects. It is, nevertheless, expected that B-type stars also maintain convective core dynamos \citep{aug2016}.
\end{itemize}

Therefore, while according to magnetic flux conservation the weakening of the polar field strength is accounted for in our models, it is not evident how these fields transform as the star crosses the HRD. Our model calculations predict that near-surface convective zones appear after the TAMS, that may give rise to dynamo activity. Such dynamo activity is indeed observed in cool massive stars \citep{grunhut2010}. 

It remains a puzzling question to what depth fossil fields, which form a large-scale magnetosphere above the stellar surface, penetrate in the stellar interiors (see e.g. \citealt{mahes1988,braithwaite2006}). To more appropriately model magnetic braking, the depth to which the magnetic torque is exerted in the stellar interior should be established. The layers with strong magnetic torque are indeed expected to maintain a nearly flat angular velocity profile, however if there exist adjacent layers where the torque has ceased, then this could introduce a jump in the angular velocity profile leading to the development of strong shears. On the other hand, the strong shears may increase the efficiency of angular momentum transport, hence mitigating the break in the angular velocity profile. This is why the assumptions regarding internal angular momentum and chemical element transport are critical, because two fundamentally different scenarios yield different model responses to surface magnetic braking. Nonetheless, it is clear that magnetic fields in the stellar interior need to be considered jointly with stellar rotation \citep[e.g.,][]{mestel1987b,maeder2003,heger2005}. Multi-dimensional approaches have been invoked to study the interaction between stellar rotation and internal magnetic fields \citep[e.g.,][]{mathis2005,duez2010,mathis2011,mathis2012}, and recent studies have begun to focus on this interaction, primarily via the magneto-rotational instability in massive and intermediate-mass stellar evolution models \citep{wheeler2015,quentin2018}.

\section{Conclusions}

We have discussed the combined effects of mass-loss quenching and magnetic braking, considering an evolving dipolar surface fossil magnetic field, in the Geneva stellar evolution code, and in this work we have shown how 15 M$_{\odot}$ single star models evolve with and without surface magnetic fields. We computed non-rotating models, rotating models neglecting $D_{\rm ST}$ in the angular momentum transport equation to allow for differential rotation, and rotating models including $D_{\rm ST}$ in the angular momentum transport equation to achieve a flattened $\Omega$ profile. 

The key results of this study are the following:
\begin{itemize}
\item We showed that in the case of 15 M$_\odot$ stellar models, mass-loss quenching due to a fossil field is modest since mass loss over the lifetime of the star is modest compared to its initial mass. However, the rotational evolution of the star models is strikingly different, even if the HRD tracks are nearly identical as is the case for M5 (RS-) and M6 (RSB).   
\item We identified that even for 15 M$_{\odot}$ models the evolutionary tracks are notably different between magnetic and non-magnetic models if differential rotation is considered (i.e., the models evolve at higher luminosity if the star hosts a magnetic field). This is primarily a consequence of the rapid and strong shear mixing induced by magnetic braking. In non-rotating but initially higher-mass models ($>$ 40 M$_\odot$)\break \cite{petit2017} found that the increase of stellar luminosity results from magnetic mass-loss quenching alone.   
\item We found that main sequence rotating models which include a surface magnetic field evolve to a region of the Hunter diagram where the anomalous Group 2 objects are located. In accord with the results of \cite{meynet2011}, magnetic braking enhances the chemical enrichment if the star undergoes radial differential rotation, however the enrichment is reduced if the star rotates as a solid body. We found that including magnetic field evolution results in weakening magnetic braking, hence our magnetic models with solid-body rotation do evolve into Group 2 unlike the models by \cite{meynet2011}. 
\item Accounting for the evolution of the surface field, we placed constraints on the observable magnetic field strength for future spectropolarimetric observations and we studied the time evolution of the magnetospheric parameters. We found that the surface polar field strength weakens at maximum by an order of magnitude on the main sequence, however as the star crosses the HRD the decrease becomes more rapid. Consequently, the initially strong magnetic confinement results in an escaping wind fraction of only 10-20\% during the early evolution of the star. As the rotation of the star evolves, the Kepler co-rotation radius increases systematically, whereas the Alfv\'en radius decreases. In its early evolution, the star transitions from the centrifugally-supported magnetosphere regime to the dynamical magnetosphere regime of the confinement-rotation diagram. 
\end{itemize}
To gain further insights towards understanding massive stars with surface fossil magnetic fields, the following points should be considered:
\begin{itemize}
\item The effects on the stellar surface by fossil magnetic fields can be incorporated into stellar evolution model calculations by means of scaling relations. However, it remains largely uncertain how the surface fields behave in the subsurface layers and throughout the radiative envelope of the star. Therefore, developing a formalism to account for that would be advantageous to constrain the angular momentum transport mechanisms used in stellar evolution model calculations.  
\item It is evident that due to the model dependence on stellar rotation, mass loss, initial mass, metallicity, and other parameters, a more comprehensive (parameter) study is required. To this extent, we are computing a large grid of models that will explore the above-mentioned parameter space in more detail, in order to study the consequences of individual parameters on model predictions (Keszthelyi et al., in prep.).  
\item Based on our model calculations, it would be worthwhile to invest in undertaking two large-scale observational surveys. i) There is interest in understanding how fossil fields evolve with time (the LIFE collaboration, \citealt{martin2018}), however data from stars at late evolutionary phases remain scarce. It would be valuable to know if evolved massive stars with late B and A spectral types retain weak, organized surface magnetic fields consistent with the characteristics and incidence of surface magnetism in the OB phase. ii) It has been proposed that Group 2 stars on the Hunter diagram may have surface magnetic fields, however no systematic survey of magnetism in these objects has been carried out. In our solar metallicity models the polar magnetic field strength weakens from a few kG to $\sim$ 800 G in the Group~2 star regime of the Hunter diagram. It would be worthwhile to investigate the fractional incidence of surface magnetism in known Group 2 stars.  
\end{itemize}

\section*{Acknowledgements}

We appreciate great discussions with R. Townsend,\break A. ud-Doula, J. Puls, and M. Shultz. We thank the anonymous referee for helpful comments on the manuscript. 
Z.K. greatly acknowledges the warm hospitality at the Geneva Observatory, where part of this research was carried out. 
G.M. and C.G. acknowledge support from the Swiss National Science Foundation (project number 200020-172505).
G.A.W. acknowledges support in the form of a Discovery Grant from the Natural Science and Engineering Research Council (NSERC) of Canada.
V.P. acknowledges support provided by 
(i) the National Aeronautics and Space Administration through Chandra Award Number GO3-14017A issued by the Chandra X-ray Observatory Center, which is operated by the Smithsonian Astrophysical Observatory for and on behalf of the National Aeronautics Space Administration under contract NAS8-03060, and
(ii) program HST-GO-13734.011-A that was provided by NASA through a grant from the Space Telescope Science Institute, which is operated by the Association of Universities for Research in Astronomy, Inc., under NASA contract NAS 5-26555.
A.D.U. acknowledges support from NSERC.
 

\bibliographystyle{mnras}
\bibliography{ref} 


\appendix

\section{Tables} 

%
%
Tables \ref{tab:t1}, \ref{tab:t2}, and \ref{tab:t3} show the stellar and magnetospheric parameters throughout the main sequence evolution of the 15~$M_{\odot}$ models with and without rotation. $X_c$ denotes the core hydrogen content.
%
%
\begin{table*}
\caption{Stellar and magnetospheric parameters of M1 (- - -) and M2 (- -B) at various evolutionary phases.}
\centering
\begin{tabular}{lcccccc}   
\hline
 &  ZAMS & 75\% X$_c$ & 50\% X$_c$ &25\% X$_c$ & TAMS &   \\
\hline \hline
Model: M1 (- - -) \\
\hline 
 $T_{\rm eff}$ [kK] & 31.705  &  30.250  &  28.805  &  26.485  &  27.005  \\ 
 $\log (L/L_{\odot})$ &4.266  &  4.367  &  4.458  &  4.537  &  4.650  \\
 age [Myr] & 0.112  &  4.645  &  7.706  &  9.807  &  11.258 \\
 $M \, [M_{\odot}]$ & 15.000  &  14.950  &  14.902  &  14.854  &  14.808  \\
 $R \, [R_{\odot}]$ & 4.544  &  5.602  &  6.863  &  8.891  &  9.746  \\
 $\log$ g [cm s$^{-1}$] & 4.299  &  4.116  &  3.938  &  3.712  &  3.631  \\
 $^1$H$_{\rm surf}$ [mass frac.]&  0.720  &  0.720  &  0.720  &  0.720  &  0.720 \\
 $^4$He$_{\rm surf}$ [mass frac.]&  0.266  &  0.266  &  0.266  &  0.266  &  0.266 \\
 $^{12}$C$_{\rm surf} $ [$\cdot 10^{-3}$] [mass frac.]&  2.283  &  2.283  &  2.283  &  2.283  &  2.283  \\
 $^{14}$N$_{\rm surf}$ [$\cdot 10^{-3}$] [mass frac.]&   0.659 & 0.659 & 0.659 & 0.659& 0.659 \\
 $^{16}$O$_{\rm surf}$ [$\cdot 10^{-3}$] [mass frac.] & 5.718  &  5.718  &  5.718  &  5.718  &  5.718 \\
\hline
 $\log$ $\dot{M}$ [M$_{\odot}$ yr$^{-1}$] & -8.047 & -7.878 & -7.718 & -7.572 & -7.356 \\
 v$_\infty$ $[\mathrm{km \, s^{-1}}]$  &   2877 &  2575 &  2312 &  2015 &  1906 \\
\hline  
\hline 
Model: M2 (- -B)\\
\hline 
 $T_{\rm eff}$ [kK] & 31.705  &  30.289  &  28.875  &  26.588  &  27.173 \\ 
 $\log (L/L_{\odot})$ & 4.266  &  4.370  &  4.464  &  4.545  &  4.660 \\
  age [Myr] &  0.112  &  4.633  &  7.679  &  9.764  &  11.200 \\
 $M \, [M_{\odot}]$ & 15.000  &  14.994  &  14.988  &  14.980  &  14.972 \\
 $R \, [R_{\odot}]$ & 4.544  &  5.609  &  6.879  &  8.908  &  9.733  \\
 $\log$ g [cm s$^{-1}$] & 4.299  &  4.116  &  3.939  &  3.714  &  3.637 \\
 $^1$H$_{\rm surf}$ [mass frac.]&  0.720  &  0.720  &  0.720  &  0.720  &  0.720 \\
 $^4$He$_{\rm surf}$ [mass frac.]&  0.266  &  0.266  &  0.266  &  0.266  &  0.266 \\
 $^{12}$C$_{\rm surf} $ [$\cdot 10^{-3}$] [mass frac.]&  2.283  &  2.283  &  2.283  &  2.283  &  2.283  \\
 $^{14}$N$_{\rm surf}$ [$\cdot 10^{-3}$] [mass frac.]&   0.659 & 0.659 & 0.659 & 0.659& 0.659 \\
 $^{16}$O$_{\rm surf}$ [$\cdot 10^{-3}$] [mass frac.] & 5.718  &  5.718  &  5.718  &  5.718  &  5.718 \\
\hline
 $\eta_\star$ & 2404  &  1174  &  597  &  288  &  158 \\
 $B_p$ [G] & 4000  &  2625  &  1745  &  1040  &  872   \\
 $R_A$ [$R_\star$] & 7.292  &  6.144  &  5.233  &  4.410  &  3.838  \\
 $f_B \, [\%]$   & 9.720  &  11.527  &  13.524  &  16.040  &  18.429 \\
 $\log$ $\dot{M}$ [M$_{\odot}$ yr$^{-1}$] &-9.059  &  -8.810  &  -8.575  &  -8.351  &  -8.072  \\
 v$_\infty$ $[\mathrm{km \, s^{-1}}]$  &  2877  &  2578  &  2317 &  2023  &  1907  \\
\hline 
\hline  
\end{tabular} \label{tab:t1}
\end{table*}
\begin{table*}
\caption{Stellar and magnetospheric parameters of M3 (R- -) and M4 (R-B) at various evolutionary phases.}
\centering
\begin{tabular}{lcccccc}   
\hline
 &  ZAMS & 75\% X$_c$ & 50\% X$_c$ &25\% X$_c$ & TAMS &   \\
\hline \hline
Model: M3 (R- -)\\
\hline 
 $T_{\rm eff}$ [kK] & 31.191  &  29.974  &  28.597  &  26.234  &  27.024\\ 
 $\log (L/L_{\odot})$ & 4.258  &  4.377  &  4.500  &  4.615  &  4.749 \\
 age [Myr] & 0.117  &  5.341  &  9.143  &  11.793  &  13.454 \\
 $M \, [M_{\odot}]$ &  15.000  &  14.941  &  14.872  &  14.791 &  14.709 \\
 $R \, [R_{\odot}]$ & 4.652  &  5.777  &  7.313  &  9.917  &  10.904\\
 $\log$ g [cm s$^{-1}$] & 4.279  &  4.089  &  3.882  &  3.615  &  3.530\\
 $^1$H$_{\rm surf}$ [mass frac.]&  0.720  &  0.720  &  0.719  &  0.715  &  0.706\\
 $^4$He$_{\rm surf}$ [mass frac.]& 0.266  &  0.266  &  0.267  &  0.271  &  0.280 \\
 $^{12}$C$_{\rm surf}$ [$\cdot 10^{-3}$] [mass frac.]& 2.283  &  2.261  &  2.004  &  1.669  &  1.415 \\
 $^{14}$N$_{\rm surf}$ [$\cdot 10^{-3}$] [mass frac.]& 0.659 & 0.682 & 1.018 & 1.571 & 2.079 \\
 $^{16}$O$_{\rm surf}$ [$\cdot 10^{-3}$] [mass frac.]& 5.718  &  5.715  &  5.631  &  5.420  &  5.170\\
 $v_{\rm rot} \, [\mathrm{km \, s^{-1}}]$ &  200.0  &  160.0  &  153.0  &  138.0  &  107.0 \\
\hline
 $\log$ $\dot{M}$ [M$_{\odot}$ yr$^{-1}$] &  -8.053  &  -7.848  &  -7.624  &  -7.405  &  -7.162 \\
 v$_\infty$ $[\mathrm{km \, s^{-1}}]$  &  2847  &  2535  &  2231  &  1893  &  1774 \\
\hline  
\hline 
Model: M4 (R-B)\\
\hline 
 $T_{\rm eff}$ [kK] &  31.200 &  31.206 &  30.579 &  28.278 &  29.102 \\
 $\log (L/L_{\odot})$ & 4.258 &  4.432 &  4.589 &  4.703 &  4.830\\
  age [Myr] &  0.117 &  7.233 &  11.650 &  13.963 &  15.285\\
 $M \, [M_{\odot}]$ &  15.000 &  14.989 &  14.975 &  14.959 &  14.941\\
 $R \, [R_{\odot}]$ &  4.650 &  5.678 &  7.080 &  9.446 &  10.319\\
 $\log$ g [cm s$^{-1}$] & 4.279 &  4.105 &  3.913 &  3.662 &  3.585 \\
 $^1$H$_{\rm surf}$ [mass frac.]&  0.720 &  0.697 &  0.674 &  0.665 &  0.663  \\
 $^4$He$_{\rm surf}$ [mass frac.]&  0.266 &  0.289 &  0.312 &  0.321 &  0.323 \\
 $^{12}$C$_{\rm surf} $ [$\cdot 10^{-3}$] [mass frac.]&  2.283 &  1.105 &  0.809 &  0.736 &  0.724 \\
 $^{14}$N$_{\rm surf}$ [$\cdot 10^{-3}$] [mass frac.]&  0.659 & 3.246 & 4.197 & 4.443 & 4.484 \\
 $^{16}$O$_{\rm surf}$ [$\cdot 10^{-3}$] [mass frac.] & 5.718 &  4.239 &  3.560 &  3.381 &  3.351\\
 $v_{\rm rot} \, [\mathrm{km \, s^{-1}}]$   & 196.0 &  14.2 &  2.1 &  0.4 &  0.2  \\
\hline
 $\eta_\star$ & 2617 & 973 & 380 & 151 & 83 \\
 $B_p$ [G] & 4000 & 2681 & 1725 & 969 & 812  \\
 $R_A$ [$R_\star$] & 7.442 & 5.875 & 4.707 & 3.797 & 3.306 \\
 $R_K$ [$R_\star$] &2.521 & 13.568 & 44.499 & 134.596 & 168.169\\
 $f_B \, [\%]$   & 9.525 & 12.053 & 15.030 & 18.627 & 21.400 \\
 $\log$ $\dot{M}$ [M$_{\odot}$ yr$^{-1}$] & -9.074 & -8.673& -8.307 & -7.999 & -7.724  \\
 v$_\infty$ $[\mathrm{km \, s^{-1}}]$  &   2855 & 2553 & 2264 & 1936 & 1818  \\
\hline 
\hline  
\end{tabular} \label{tab:t2}
\end{table*}
\begin{table*}
\caption{Stellar and magnetospheric parameters of the M5 (RS-) and M6 (RSB) at various evolutionary phases.}
\centering
\begin{tabular}{lcccccc}   
\hline
 &  ZAMS & 75\% X$_c$ & 50\% X$_c$ &25\% X$_c$ & TAMS &  \\
\hline \hline
Model: M5 (RS-)\\
\hline 
 $T_{\rm eff}$ [kK] & 31.193 & 30.442 & 29.439 & 27.022 & 27.311  \\ 
 $\log (L/L_{\odot})$ & 4.258 & 4.415 & 4.544 & 4.666 & 4.803  \\
  age [Myr] & 0.117 & 6.757 & 10.522 & 12.959 & 14.459\\
 $M \, [M_{\odot}]$ & 15.000 & 14.918 & 14.835 & 14.744 & 14.649 \\
 $R \, [R_{\odot}]$ & 4.651 & 5.847 & 7.258 & 9.912 & 11.364  \\
 $\log$ g [cm s$^{-1}$] & 4.279 & 4.078 & 3.888 & 3.614 & 3.493  \\
 $^1$H$_{\rm surf}$ [mass frac.] &  0.720 & 0.715 & 0.699 & 0.684 & 0.676  \\
 $^4$He$_{\rm surf}$ [mass frac.] &  0.266 & 0.271 & 0.287 & 0.302 & 0.310 \\
 $^{12}$C$_{\rm surf}$ [$\cdot 10^{-3}$] [mass frac.] &   2.283 & 1.864 & 1.359 & 1.110 & 1.004 \\
 $^{14}$N$_{\rm surf}$ [$\cdot 10^{-3}$] [mass frac.] &   0.659 & 1.491 & 2.806 & 3.516 & 3.822 \\
 $^{16}$O$_{\rm surf}$ [$\cdot 10^{-3}$] [mass frac.] &  5.718 & 5.288 & 4.445 & 3.966 &3.759 \\
 $v_{\rm rot} \, [\mathrm{km \, s^{-1}}]$   & 200.0 & 192.0& 185.0 & 169.0 & 177.0 \\
\hline
 $\log$ $\dot{M}$ [M$_{\odot}$ yr$^{-1}$] & -8.053 & -7.771 & -7.537 & -7.304 & -7.047\\
 v$_\infty$ $[\mathrm{km \, s^{-1}}]$  &  2847 & 2514 & 2231 & 1882& 1713  \\
\hline  
\hline 
Model: M6 (RSB)\\
\hline 
 $T_{\rm eff}$ [kK] & 31.237 & 30.594 &29.382 &26.936 &27.466 \\ 
 $\log (L/L_{\odot})$ & 4.249 & 4.407 & 4.548 & 4.667 & 4.799 \\
  age [Myr] & 0.117 & 6.159 & 10.255 & 12.829 & 14.359 \\
 $M \, [M_{\odot}]$ & 15.000 & 14.991 & 14.980 & 14.965 & 14.946 \\
 $R \, [R_{\odot}]$ & 5.660 & 5.740 & 7.309 & 9.975 & 11.184  \\
 $\log$ g [cm s$^{-1}$] & 4.108 & 4.096 & 3.886 & 3.615 & 3.515  \\
 $^1$H$_{\rm surf}$ [mass frac.]&   0.720 & 0.716 & 0.712 & 0.707 & 0.703  \\
 $^4$He$_{\rm surf}$ [mass frac.]&   0.266 & 0.270 & 0.274 & 0.279 & 0.283 \\
 $^{12}$C$_{\rm surf} $ [$\cdot 10^{-3}$] [mass frac.]&   2.283 & 1.816 & 1.589 & 1.456 & 1.389  \\
 $^{14}$N$_{\rm surf}$ [$\cdot 10^{-3}$] [mass frac.]&   0.659 & 1.517 & 2.041 & 2.374 & 2.552 \\
 $^{16}$O$_{\rm surf}$ [$\cdot 10^{-3}$] [mass frac.] & 5.718 & 5.310 & 5.002 & 4.795 & 4.680 \\
 $v_{\rm rot} \, [\mathrm{km \, s^{-1}}]$   & 189.0 & 20.6 & 6.8 & 4.5 & 2.8 \\
\hline
 $\eta_\star$ & 2607 & 1048 & 417 & 158 & 79 \\
 $B_p$ [G] & 4000 & 2599 & 1603 & 863 & 688   \\
 $R_A$ [$R_\star$] & 7.436 & 5.981 & 4.8094& 3.835 & 3.275   \\
 $R_K$ [$R_\star$] &2.585 & 10.541 & 19.632 & 25.068 & 31.995\\
 $f_B \, [\%]$   & 9.533 & 11.841 & 14.711 & 18.442 & 21.607 \\
 $\log$ $\dot{M}$ [M$_{\odot}$ yr$^{-1}$] & -9.075 & -8.724 & -8.382 & -8.060 & -7.752  \\
 v$_\infty$ $[\mathrm{km \, s^{-1}}]$  &   2852 & 2541 & 2231 & 1890 & 1755  \\
\hline 
\hline  
\end{tabular} \label{tab:t3}
\end{table*}
%
%


\bsp	
\label{lastpage}
\end{document}